\documentclass[12pt,a4paper]{article}

\usepackage{epsfig}
\usepackage{psfig}
\usepackage{exscale}
\def\lsim{\raise0.3ex\hbox{$<$\kern-0.75em\raise-1.1ex\hbox{$\sim$}}}
\def\gsim{\raise0.3ex\hbox{$>$\kern-0.75em\raise-1.1ex\hbox{$\sim$}}}
\newcommand{\be}{\begin{equation}}
\newcommand{\ee}{\end{equation}}
\newcommand{\ba}{\begin{eqnarray}}
\newcommand{\ea}{\end{eqnarray}}

\def\spose#1{\hbox to 0pt{#1\hss}}
\def\ltapprox{\mathrel{\spose{\lower 3pt\hbox{$\mathchar"218$}}
 \raise 2.0pt\hbox{$\mathchar"13C$}}}
\def\gtapprox{\mathrel{\spose{\lower 3pt\hbox{$\mathchar"218$}}
 \raise 2.0pt\hbox{$\mathchar"13E$}}}

\def\NT{N_\tau}
\def\nt{\ifmmode\NT\else$\NT$\fi}
\def\NS{N_\sigma}
\def\ns{\ifmmode\NS\else$\NS$\fi}

\def\PRep{{ Phys.\ Rep.\ }}

\def\p{^\prime}

\def\n{\noindent}

\begin{document}
\begin{titlepage} 
\thispagestyle{empty}

 \mbox{} \hfill BI-TP 2002/11\\
 \mbox{} \hfill November 2002\\
 \mbox{} \hfill cond-mat/0209492v2
\begin{center}
\vspace*{0.8cm}
{{\Large \bf Numerical equation of state and other\\
scaling functions from an improved \\
three-dimensional Ising model\\}}\vspace*{1.0cm}
{\large J. Engels, L. Fromme and M. Seniuch}\\ \vspace*{0.8cm}
\centerline {{\em Fakult\"at f\"ur Physik, 
    Universit\"at Bielefeld, D-33615 Bielefeld, Germany}} \vspace*{0.4cm}
\protect\date \\ \vspace*{0.9cm}
{\bf   Abstract   \\ } \end{center} \indent
We study an improved three-dimensional Ising model with external magnetic
field near the critical point by Monte Carlo simulations. From our data we
determine numerically the universal scaling functions of the magnetization,
that is the equation of state, of the susceptibility and of the correlation
length. In order to normalize the scaling functions we calculate the 
critical amplitudes of the three observables on the critical line, the
phase boundary and the critical isochore. The amplitudes lead to the
universal ratios $C^+/C^-=4.756(28)$, $R_{\chi}=1.723(13)$, $Q_c=0.326(3)$
and $Q_2=1.201(10)$. We find excellent agreement of the data with the
parametric representation of the asymptotic equation of state as found by
field theory methods. The comparison of the susceptibility data to the
corresponding scaling function shows a marginal difference in the symmetric
phase, which can be explained by the slightly different value for 
$R_{\chi}$ used in the parametrization. The shape of the 
correlation-length-scaling function is similar to the one of the
susceptibility, as expected from earlier parametrizations. The peak
positions of the two scaling functions are coinciding within the error bars.

\vfill \begin{flushleft} 
PACS : 64.10.+h; 75.10.Hk; 05.50+q \\ 
Keywords: Ising model; Equation of state; Scaling function; 
Correlation length; Universal amplitude ratios\\ 
\noindent{\rule[-.3cm]{5cm}{.02cm}} \\
\vspace*{0.2cm} 
E-mail: engels, fromme, seniuch@physik.uni-bielefeld.de\
\end{flushleft} 
\end{titlepage}


\section{Introduction}

The Ising model is the simplest non-trivial spin model with critical
behaviour. In three dimensions the universality class of the Ising model
includes as well many real physical systems as also theoretical models
which exhibit a continuous phase transition. Important examples are known
from condensed matter physics and high-energy physics (see for instance the 
detailed review in Ref.\ \cite{Pelissetto:2000ek}). Due to their simplicity
and the many physically relevant applications the $3d$ Ising model and 
members of the corresponding universality class belong to the most 
extensively studied systems. In these investigations both a variety of
analytic methods \cite{Pelissetto:2000ek}-\cite{Zinn-Justin:2001bf} and
Monte Carlo simulations have been used to calculate the universal quantities
characterizing the universality class, such as critical exponents and
amplitude ratios. 

In the vicinity of the critical point the behaviour of the order parameter,
the magnetization,  and the correlation length are described by
asymptotic scaling functions. The functions depend on the temperature
and on the magnetic field and are universal apart from metric factors
which are specific for the model under consideration. They are asymptotic
in so far as they are valid in the thermodynamic limit, that is for
$V\rightarrow \infty$, and for $T\rightarrow T_c$, so that irrelevant 
scaling fields can be neglected. In order to be comparable to other model
systems the scaling functions have to be normalized which amounts to fixing
the metric factors. There exist numerous parametrizations (see e.\ g.\ Refs.\
\cite{Pelissetto:2000ek} and \cite{Guida:1998bx}-\cite{Campostrini:1999at})
of the equation of state, the scaling function of the order parameter,
and other scaling functions. Since a parametric form of a scaling 
function has to describe the critical behaviour of the respective 
observable in accord with the known critical exponents and amplitude
ratios these are usually taken either as an input to fix the parameters
or as a check. Further critical quantities may then eventually be 
calculated from the parametrizations.

Instead of relying only on the known critical parameters one can in
principle directly determine the scaling functions from simulation data,
taking at most the critical exponents as input. This has
already been achieved in the case of the three-dimensional $O(2)$ 
\cite{Engels:2000xw} and $O(4)$ vector models \cite{Toussaint:1996qr,
Engels:1999wf} for the equation of state. In the $3d$ Ising model class
(the $O(1)$ vector model) no such attempt has been undertaken as far as we 
know. In this paper our main intention is to accomplish the same for 
that universality class and to compare the parametric representation of
the equation of state as found from field theory
\cite{Zinn-Justin:2001bf}-\cite{Guida:1996ep} directly to simulation
results. A further aim is to find numerically the scaling function of
the correlation length. 
It is clear, that for this project primarily simulations with non-zero
magnetic field are required. The usual simulations at zero magnetic field
can at best serve to normalize the scaling functions. The principal
difficulties which we expect are due to the use of finite volumes in
the simulations and to the finiteness of the critical region where
scaling works. In order to attenuate these problems we use relatively
large lattices with up to 120 points in each direction and we employ an
improved Ising model to reduce corrections to scaling.

The model which we want to study is the one-component $\phi^4$ or
Landau-Ginzburg model. It is defined by 
\be
\beta\,{\cal H}\;=\;-J \,\sum_{<x,y>} \phi_x \phi_y
       \;+\;\sum_{x}[\phi_x^2 +\lambda (\phi_x^2-1)^2]  
       \;-\; H \,\sum_{x} \phi_x \;.
\label{act}
\ee
Here $x$ and $y$ are the nearest-neighbour sites on a three-dimensional 
simple cubic lattice, $\phi_x$ is the field variable at site $x$ and $H$
is the external magnetic field. We consider the coupling constant $J$ 
as inverse temperature, that is $J=1/T$. In contrast to the ordinary 
Ising model the length of the spins $\phi_x$ in Eq.\ (\ref{act})
is not fixed to one. For $\lambda=0$ one obtains the Gaussian model,
whereas in the limit $\lambda \rightarrow \infty$ the simple Ising
model is recovered. By choosing an appropriate $\lambda$-value it is
then possible to eliminate leading order corrections to scaling. This 
has been shown numerically in Refs.\ 
\cite{Ballesteros:1998my}-\cite{Hasenbusch:1999mw} for $H=0$. 
There is however a drawback in using a finite $\lambda\:$: we cannot 
directly verify our results by comparison with the known, non-universal
amplitudes (see e.\ g.\ Ref.\ \cite{Zinn:1996sy}) of the normal Ising
model.

The observables which we want to measure in our simulations are the
magnetization $M$, the susceptibility $\chi$ and the correlation
length $\xi$. The magnetization is the expectation value of the 
lattice average $\phi$ of the spin variable
\be
M \;=\; \langle\: \frac{1}{V}\sum_{x} \phi_x\: \rangle\;
 =\; \langle \,  \phi\, \rangle~,
\label{truem}
\ee
where $V=L^3$ and $L$ is the number of lattice points per direction.
As long as $H$ is non-zero and the volume large enough there is no 
problem with this definition. At zero magnetic field, $H=0$, however,
the lattice average of the spins will have a vanishing expectation 
value on all finite lattices, that is $\langle \: \phi \: \rangle = 0\,$.
At very small magnetic fields and/or small $L$ we approximate $M$
therefore by \cite{Talapov:1996yh} 
\be
M \;\simeq \; \langle |\phi|\, \rangle~.
\label{magmod}
\ee
On finite lattices the magnetization of Eq.\ (\ref{magmod})
approaches the infinite volume limit from above, whereas $M$
as defined by Eq.\ (\ref{truem}) for $H\ne 0$ reaches the 
thermodynamic limit from below. The susceptibility is defined as 
usual by the derivative of the magnetization
\be
\chi = {\partial M \over \partial H}
 \;=\; V(\langle \: \phi^2\: \rangle -M^2)~. 
\label{chi}
\ee
For the determination of the correlation length we consider three 
different definitions. The second moment correlation length
is obtained from 
\be
\xi_{2nd} \;=\; \left( {\mu_2 \over 2d \mu_0} \right)^{1/2}~,
\label{xi2nd}
\ee
where the $\mu_n$ are the moments of the connected part of the
correlation function
\ba
\mu_n \!\!&=&\!\! \sum_{x} x^n \langle \: \phi_x \phi_0 \: \rangle_c~,  
\label{mome} \\
\langle \: \phi_x \phi_0 \: \rangle_c  \!\!&=&\!\! 
\langle \: \phi_x \phi_0 \: \rangle -\langle \,  \phi\, \rangle^2~,
\label{corc}
\ea 
and $x^2= {\bf x}^2$. The moment $\mu_0$ is 
equal to the susceptibility. A more and more often used definition 
of the correlation length is
\be
\xi^F \;=\; \left( {\chi / F -1 \over 4 \sin^2(\pi/L)} \right)^{1/2}~,
\label{xif}
\ee
where $F$ is the Fourier transform of the correlation function at
momentum $p_{\mu}=2\pi {\hat e}_{\mu}/L$, and ${\hat e}_{\mu}$ a unit
vector in one of the three directions
\be
F \;=\; {1 \over V} \langle | \sum_{x}\exp (ip_{\mu}x) \phi_x |^2
\rangle~.
\label{eff}
\ee
In the simulations we compute $F$ from an average over all three 
directions. Finally, the exponential or "true" correlation length 
$\xi_{exp}$ governs the large distance behaviour of the connected
part of the correlation function
\be
\langle \: \phi_x \phi_0 \: \rangle_c \sim 
\exp (-|{\bf x}|/\xi_{exp})~.
\label{xiexp}
\ee 
We explain later in detail how we determine this observable from
the correlation function data.

The rest of the paper is organized as follows. First we discuss the
critical behaviour of the observables and the universal scaling 
functions, which we want to calculate. In Section 3 we describe 
some details of our simulations. Section 4 serves to determine 
the amplitudes of the magnetization and the susceptibility. In 
the following Section 5 we compare our data for the equation of state
to a parametrization from field theory. The determination of the 
correlation length and its amplitudes is described in Section 6. 
In Section 7 we present our results for the scaling function of the
correlation length. We close with a summary and the conclusions.


\section{Critical Behaviour and Scaling Functions}
\label{section:Criti}


In the thermodynamic limit ($V\rightarrow \infty$) the observables show
power law behaviour close to $T_c$. It is described by critical amplitudes
and exponents of the reduced temperature $t=(T-T_c)/T_c$. The scaling laws
at $H=0$ are for:

\n the magnetization 
\be
 M  \;=\; B (-t)^{\beta} \quad {\rm for~} t<0~,
\label{mcr}
\ee
the susceptibility
\be
 \chi \;=\; C^{\pm} |t|^{-\gamma} \quad {\rm for~} t \rightarrow \pm 0~,
\label{chicr}
\ee 
and the correlation length (without fixing for the moment a special 
definition) 
\be
 \xi \;=\; \xi^{\pm} |t|^{-\nu} \quad {\rm for~} t \rightarrow \pm 0~.
\label{xicr}
\ee
On the critical line $T=T_c$ or $t=0$ we have for $H>0$ 
the scaling laws
\be
M \;=\; B^cH^{1/\delta} \quad {\rm or}\quad H \;=\;D_c M^{\delta}~,
\label{mcrh}
\ee
\be
\chi \;=\; C^cH^{1/\delta-1} \quad {\rm with}\quad C^c \;=\;B^c/\delta~,
\label{chicrh}
\ee
and for the correlation length 
\be
\xi \;=\; \xi^c H^{-\nu_c}~,\quad \nu_c\;=\; \nu /\beta\delta~.
\label{xicrh}
\ee
We assume the following hyperscaling relations among the
critical exponents to be valid
\be
2-\alpha  \;=\; d\nu, \quad \gamma \;=\; \beta (\delta -1), \quad
d\nu \;=\; \beta (1 +\delta)~.
\label{hyps}
\ee
As a consequence only two critical exponents are independent. Because of
the hyperscaling relations and the already implicitly assumed equality
of the critical exponents above and below $T_c$ one can construct a
multitude of universal amplitude ratios \cite{Privman:1991} (see also the 
discussion in Ref.\ \cite{Pelissetto:2000ek}). The following list of
ratios contains those which we will determine here
\ba
&\!\!\!\! \!\! U_2  =\; C^+/C^-~, \quad &U_{\xi} 
\; = \; \xi^+ /\xi^- ~, \label{uratios}\\
&\; R_{\chi} \; =\; C^+ D_c B^{\delta-1}~, \quad &Q_c
\; =\; B^2 (\xi^+)^d /C^+ ~,\label{rce}
\ea
and 
\be
Q_2 \; =\; (\xi^c /\xi^+)^{\gamma/\nu} C^+/ C^c~.
\label{hratios}
\ee 

The critical behaviours of the magnetization and the susceptibility 
originate from the singular part of the free energy density. In the 
thermodynamic limit it obeys the scaling law  
\be
f_s( u_t,u_h) \; =\; b^{-d} f_s(b^{y_t} u_t, b^{y_h}u_h)~.
\label{freee}
\ee 
Here $b$ is a free length rescaling factor, and
\be
 y_t \; =\; 1/ \nu~, \quad \quad y_h\; =\; 1/ \nu_c~.
\label{whys}
\ee 
We have neglected possible dependencies on irrelevant scaling fields 
$u_i$ with exponents $y_i <0$. Close to $T_c$ and for 
$H \rightarrow 0$ the remaining relevant scaling fields $u_t$ and $u_h$ 
are proportional to the reduced temperature, $u_t \sim t$, and
the magnetic field, $u_h \sim H$, that is one can replace $u_t$ and
$u_h$ by a normalized reduced temperature $\bar t= tT_c/T_0$
and magnetic field $h=H/H_0$. To be definite we assume in the following
that $H$ is positive. Choosing the scale factor $b$ such that $b^{y_h}h=1$
we find
\be
f_s \; =\; h^{d\nu_c} \Phi_s( \bar t/h^{1/\beta\delta})~,
\label{fscale}
\ee 
and using $M =-\partial f_s/ \partial H$ we arrive at
\be
M \; =\; h^{1/\delta} f_G( \bar t/h^{1/\beta\delta})~,
\label{mscale}
\ee 
where $\Phi_s$ and $f_G$ are universal scaling functions once the 
normalizations are fixed. The relation (\ref{mscale}) is one form of the 
magnetic equation of state. Alternatively, one may use the Widom-Griffiths
form of the equation of state \cite{Widom:1965,Griffiths:1967}
\be
y \; =\; f(x)~,
\label{wigr}
\ee
where
\be
y \equiv h/M^{\delta}~, \quad x \equiv \bar t/M^{1/\beta}~.
\label{xandy}
\ee
We take the standard normalization conditions
\be
f(0)\; =\; 1~, \quad f(-1)\; =\; 0~,
\label{normf}
\ee
which imply
\be
M(t=0) = h^{1/\delta} \quad {\rm and } \quad H_0 = D_c~,
\label{normh}
\ee
\be
M(h=0) = (-\bar t\;)^{\beta} \quad {\rm and } \quad T_0 = B^{-1/\beta}T_c~.
\label{normt}
\ee
The two forms (\ref{mscale}) and (\ref{wigr}) are of course equivalent. The
function $f_G(z)$ and its argument $z$ are related to $x$ and $y$ by
\be
f_G = y^{-1/\delta}~, \quad z \equiv \bar t/h^{1/\beta\delta}
= xy^{-1/\beta\delta}~.
\label{xyzf}
\ee
Correspondingly the normalization conditions (\ref{normf}) translate into
\be
f_G(0)\; =\; 1~,\quad {\rm and}\quad f_G(z) {\raisebox{-1ex}{$
\stackrel{\displaystyle\longrightarrow}{\scriptstyle z \rightarrow -\infty}$}}
(-z)^{\beta}~.
\label{normfg}
\ee
Since the susceptibility $\chi$ is the derivative of $M$ with respect to $H$ 
we obtain from Eq.\ (\ref{mscale})
\be
\chi = {\partial M \over \partial H} = {h^{1/\delta -1} \over H_0} f_{\chi}(z)
~,
\label{cscale}
\ee
with 
\be
f_{\chi}(z) = {1 \over \delta} \left( f_G(z) - {z \over \beta} f_G\p (z)
\right)~.
\label{fchi}
\ee
The asymptotic behaviour of $f_{\chi}$ for $H\to 0$ at fixed $t$, that is 
for $z\rightarrow \pm \infty$, is determined by Eq.\ (\ref{chicr})
\be
f_{\chi} (z)\; {\raisebox{-1ex}{$\stackrel 
{\displaystyle =}{\scriptstyle z \rightarrow \pm\infty}$}}
\;  C^{\pm} D_c B^{\delta-1} (\pm z)^{-\gamma}~.
\label{fcasy}
\ee
We note, that the amplitude for $z\rightarrow \infty$ is simply the universal
ratio $R_{\chi}$, whereas for $z\rightarrow -\infty$ it is the analogous
quantity $R_{\chi}^- = R_{\chi}/U_2$. The leading terms for $f_G$ are
respectively
\ba
f_G (z)\!\!& {\raisebox{-1ex}{$\stackrel 
{\displaystyle =}{\scriptstyle z \rightarrow -\infty}$}}
&\!\! (-z)^{\beta} + R_{\chi}^- (- z)^{-\gamma}~,\label{fgasym}\\
f_G (z)\!\!& {\raisebox{-1ex}{$\stackrel 
{\displaystyle =}{\scriptstyle z \rightarrow +\infty}$}}
&\!\! R_{\chi} z^{-\gamma}~,
\label{fgasyp}
\ea
in accord with the normalization (\ref{normfg}). The asymptotical identity 
$f_G = f_{\chi}$ for $z\to \infty$ is due to the fact that for $T>T_c$
and small magnetic field $M$ is proportional to $H$.

The correlation length fulfills a scaling law analogous to the one of the
free energy density
\be
\xi( u_t,u_h) \; =\; b\; \xi(b^{y_t} u_t, b^{y_h}u_h)~.
\label{xisl}
\ee
In the same way as before we obtain from this equation the dependence 
of the correlation length on $\bar t$ and $h$ in the critical region
and the thermodynamic limit
\be 
\xi  \; =\; h^{-\nu_c} g_{\xi} (z)~,
\label{xiscale}
\ee
in terms of a scaling function $g_{\xi} (z)$. Here, we have assumed that
the metric factors in the scaling functions for $f_s$ and $\xi$ are the 
same, which is the usual assumption \cite{Privman:1991}.
The function $g_{\xi} (z)$ is then universal except for one normalization
factor. On the critical line $t=0$ or $z=0$ we find from (\ref{xicrh})
\be
g_{\xi} (0) \; =\; \xi^c D_c^{-\nu_c}\; =\;\xi^c (B^c)^{\nu/\beta}~,
\label{gxi0}
\ee
and from (\ref{xicr}) the asymptotic behaviour at $z \to \pm \infty$
\be
g_{\xi} (z) \; {\raisebox{-1ex}{$\stackrel 
{\displaystyle =}{\scriptstyle z \rightarrow \pm\infty}$}} \;
\xi^{\pm} B^{\nu/\beta} (\pm z)^{-\nu}~.
\label{gxiasy}
\ee
Indeed, the ratio of the amplitude for $z\to \infty$ in (\ref{gxiasy}) 
and $g_{\xi} (0)$ is universal
\be
{\xi^{+} B^{\nu/\beta} \over \xi^c (B^c)^{\nu/\beta}} \; =\; 
\left({ \delta R_{\chi} \over Q_2} \right)^{\nu/\gamma} ~, 
\label{univers}
\ee
whereas $g_{\xi} (0)$ itself is not universal.
Pelissetto and Vicari \cite{Pelissetto:2000ek} associate a universal 
scaling function $f_{\xi}(x)$ to the correlation length by the relation 
\be
\xi^2 \; =\; (B^c)^{2\delta /d} M^{-2\nu /\beta} f_{\xi}(x)~,
\label{efxi}
\ee
where $x$ is defined as in (\ref{xandy}). It is straightforward to show
that
\be
{f_{\xi}(x) \over f_{\xi}(0)} \; =\; \left( {g_{\xi}(z) \over g_{\xi}(0)}
\cdot f_G(z)^{\nu /\beta} \right)^2~.
\label{fxig}
\ee
The value $x=0$ corresponds to $z=0$. Since both $f_{\xi}(x)$ and
$f_G(z)$ are universal this is also true for
\be
{\hat g}_{\xi}(z) \; = \; g_{\xi}(z)/g_{\xi}(0)~.
\label{gnorm}
\ee
As we shall see, the scaling functions of the correlation length and of
the susceptibility have similar shapes, each with a peak at a positive
$z$-value. It is therefore useful to consider the ratio $\xi^2/\chi$.
We can express it with the corresponding ratio of universal scaling
functions
\be
{\xi^2 \over \chi} \; = \; (\xi^c)^2 (B^c)^{-1} H^{-\eta\nu_c}
\cdot {{\hat g}_{\xi}^2(z) \over f_{\chi}(z)}~.
\label{rxichi}
\ee   
Here $\eta= 2-\gamma/\nu\;$ is the exponent describing the behaviour of
the two-point function at the critical point. We postpone the discussion
of parametrizations of the scaling functions to the sections where we
compare them to our data.

\clearpage

\section{Simulation Details}
\label{section:update}

\n In the numerical simulation of the model (\ref{act}) we followed the
examples given by Brower and Tamayo \cite{Brower:mt} and Hasenbusch 
\cite{Hasenbusch:1999mw}. They combined a simple local Metropolis algorithm
for the update of the field with a cluster algorithm, which only updates
the sign of the fields $\phi_x$ at fixed values of the modulus $|\phi_x|$.
During the cluster update the system is therefore treated as an Ising
model with link-dependent coupling constants $J_{xy}=J|\phi_x||\phi_y|$
and local magnetic fields $H_x=H|\phi_x|$. In order to achieve ergodicity
of the entire update process the two algorithms have to be used in an
alternating manner. In a local Metropolis step a new proposal $\phi_x\p$
for the field is produced by \cite{Hasenbusch:1999mw} 
\be
\phi_x\p \; =\; \phi_x +s\left( r -{1 \over 2}\right)~,
\label{metro}
\ee
where $r$ is a uniformly in (0,1] distributed random number and $s$ an
adjustable size parameter. The prescription (\ref{metro}) obviously allows 
for changes of the lengths of the spins. For the cluster algorithm
we chose Wolff's single cluster update \cite{Wolff:1988uh}, which was 
modified to include a non-zero magnetic field \cite{Dimitrovic:yd}. This
variant was also successfully used in previous simulations of the
$O(2)$ \cite{Engels:2000xw} and $O(4)$ spin models \cite{Engels:1999wf}.
A complete Monte Carlo update consists then of $n$ consecutive cluster
updates followed by a Metropolis sweep of the entire lattice.
  
As explained in the introduction we aim at minimizing the corrections to
scaling. We have therefore fixed the parameter $\lambda$ of the model to
the optimal value $\lambda=1.1$ found by Hasenbusch \cite{Hasenbusch:1999mw}.    
The very precise value $J_c=0.3750966(4)$ for the corresponding critical
coupling was adopted from the same paper. All our simulations were done 
on three-dimensional lattices with periodic boundary conditions and
linear extensions ranging from $L=24$ to $L=120$. The 
size parameter $s$ in Eq.\ (\ref{metro}) was set to $s=2$. That 
resulted in acceptance rates of $59\%$ to $67\%$ in the Metropolis step
for the simulated $J$ and $H$-values. The number $n$ of cluster updates
was chosen in the region $10 \le n \le 25$ for $T>T_c$ and in the cold 
phase ($T<T_c$) in the range $3\le n \le 8$. Between two measurements
of the observables 20 to 50 Monte Carlo updates were performed, such
that the integrated autocorrelation times of the energy and the 
magnetization were $\tau_{int}\ltapprox 3$ for $L\le 48$ and 
$\tau_{int}\ltapprox 5$ for $L\ge 72$. In general we made 20000
measurements, for $L=24$ sometimes 40000. 

The coupling constant region which we have explored was $0.365 \le J 
\le 0.4$, the magnetic field was varied from $H=0$ to $H=0.005$. The
critical region, where asymptotic scaling works, is clearly inside this
parameter range. We come back to this point when we discuss our data for
the scaling functions. Further details on the Monte Carlo update, for 
example on the distribution of the modulus of the spins, can be found in
Ref.\ \cite{Seniuch:2002}.  

\section{The Critical Amplitudes of $M$ and $\chi$}
\label{section:mandchi}

\begin{figure}[b]
\begin{center}
   \epsfig{bbllx=63,bblly=280,bburx=516,bbury=563,
       file=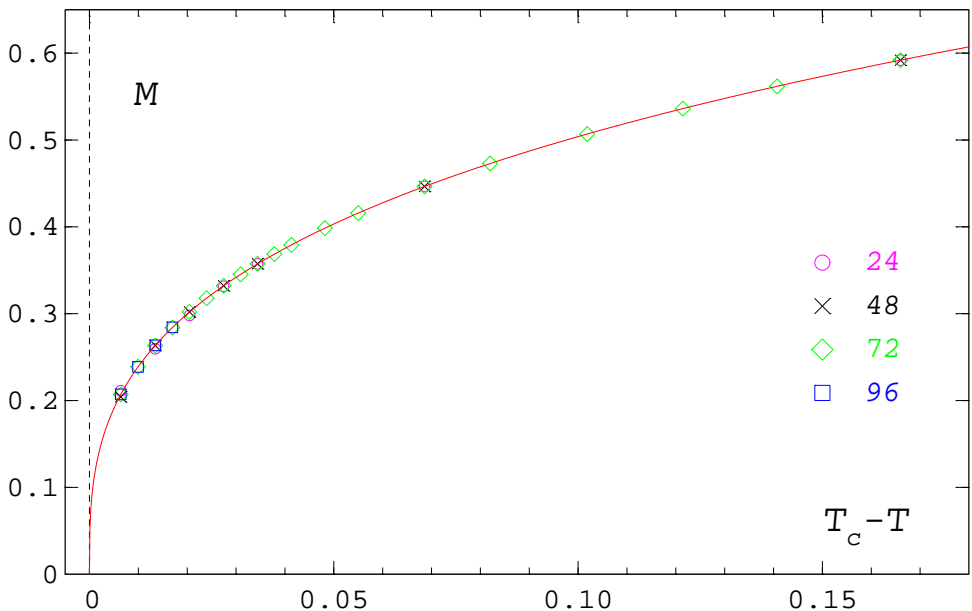, width=120mm}
\end{center}
\caption{The magnetization $M$ at $H=0$ versus $T_c-T$ for $L=24,48,72$
and 96. The solid line shows the fit (\ref{bestmt}).}
\label{fig:mhzero}
\end{figure}

\n We consider first our results for the magnetization at $H=0$ and $T<T_c$.
The primary objective here is to determine the normalization $T_0$ or
equivalently the critical amplitude $B$ of (\ref{mcr}). In Fig.\
\ref{fig:mhzero} we show $M$ from Eq.\ (\ref{magmod}) as a function of 
$T_c-T$ for different lattice sizes. The temperatures correspond to 
coupling constants in the range $0.376\le J\le 0.4\,$. In order to 
eliminate finite-size effects at a fixed $J$ (or $T$), we have simulated 
at increasingly larger values of $L$.

We have fitted the results from the largest lattices to the following ansatz
taking into account possible analytic and non-analytic corrections to scaling 
\be
M\; = \; b_0 (T_c-T)^{\beta} [1 +b_1 (T_c-T)^{\omega\nu} +b_2 (T_c-T)]~.
\label{mcorr}
\ee
\n For the fits we used the critical exponents of 
Ref.\ \cite{Zinn-Justin:2001bf} as input \hfill 
\be
\beta = 0.3258(14)~,\quad \nu = 0.6304(13)~,\quad \omega = 0.799(11)~,
\label{critex}
\ee
because we wanted to test the corresponding equation of state. The 
remaining ex\-po\-nents were determined from the hyperscaling relations
(\ref{hyps}). Instead of these field theory results we could as well
have used the exponents of Hasenbusch \cite{Hasenbusch:1999mw} or the
estimates proposed by Pelissetto and Vicari \cite{Pelissetto:2000ek}.
The difference is marginal, the various exponent results coincide within
the error bars. The fit with the ansatz (\ref{mcorr}) nicely confirms the
expected absence of the leading non-analytic correction term: $b_1$ is zero
within error bars. The best fit to the points with $T_c-T \:\lsim \: 0.10$
and $b_1\equiv 0$ is given by
\be
M\; = \; 1.0735(2)(T_c-T)^{\beta} [1 - 0.061(2)(T_c-T)]~,
\label{bestmt}
\ee
with $\chi^2/N_f=0.8\,$. From $b_0$ we obtain 
\be
T_0\: = \: b_0^{-1/\beta}\: = \: 0.8044(4)~,\quad B\: =\:b_0 T_c^{\beta}
\: =\: 1.4776(2)~.
\label{tee0}
\ee
\begin{figure}[t!]
\begin{center}
   \epsfig{bbllx=63,bblly=280,bburx=516,bbury=563,
       file=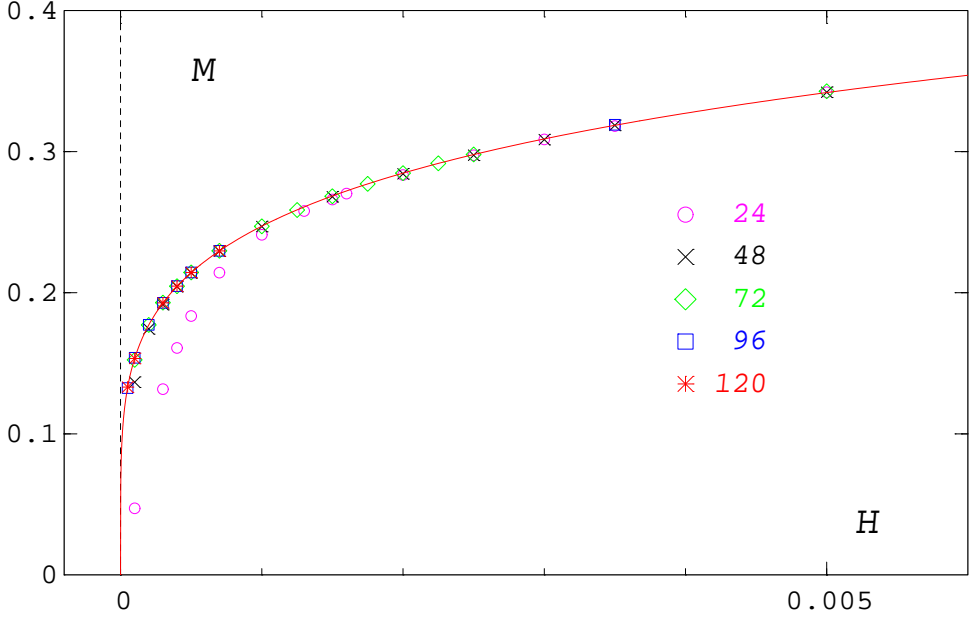, width=120mm}
\end{center}
\caption{The magnetization $M$ at $T_c$ as a function of the magnetic field
$H$ for $L=24,48,72,96$ and 120. The solid line shows the fit (\ref{bestmh}).}
\label{fig:mtzero}
\end{figure}
\indent On the critical isotherm, that is at $T_c$, we have measured the 
magnetization for $H>0$ on lattices with $L=24,48,72,96$ and 120 using
Eq.\ (\ref{truem}). The results are shown in Fig.\ \ref{fig:mtzero}. We
observe an increasing finite size dependence close to $H=0$. The results
from the largest lattices have been fitted to the ansatz
\be
M\; =\; B^c H^{1/\delta} [1 + B^c_1 H^{\omega\nu_c} + B^c_2 H]~.
\label{mhcorr}
\ee
In the $H$-interval $[0.00005,\;0.0005]$, however, only the leading term is
of relevance. A fit in this interval with simply the first term in 
(\ref{mhcorr}) leads to
\be 
B^c \; = \; 1.0435(15)~.
\label{bece}
\ee   
\begin{figure}[t!]
\begin{center}
   \epsfig{bbllx=63,bblly=280,bburx=516,bbury=563,
       file=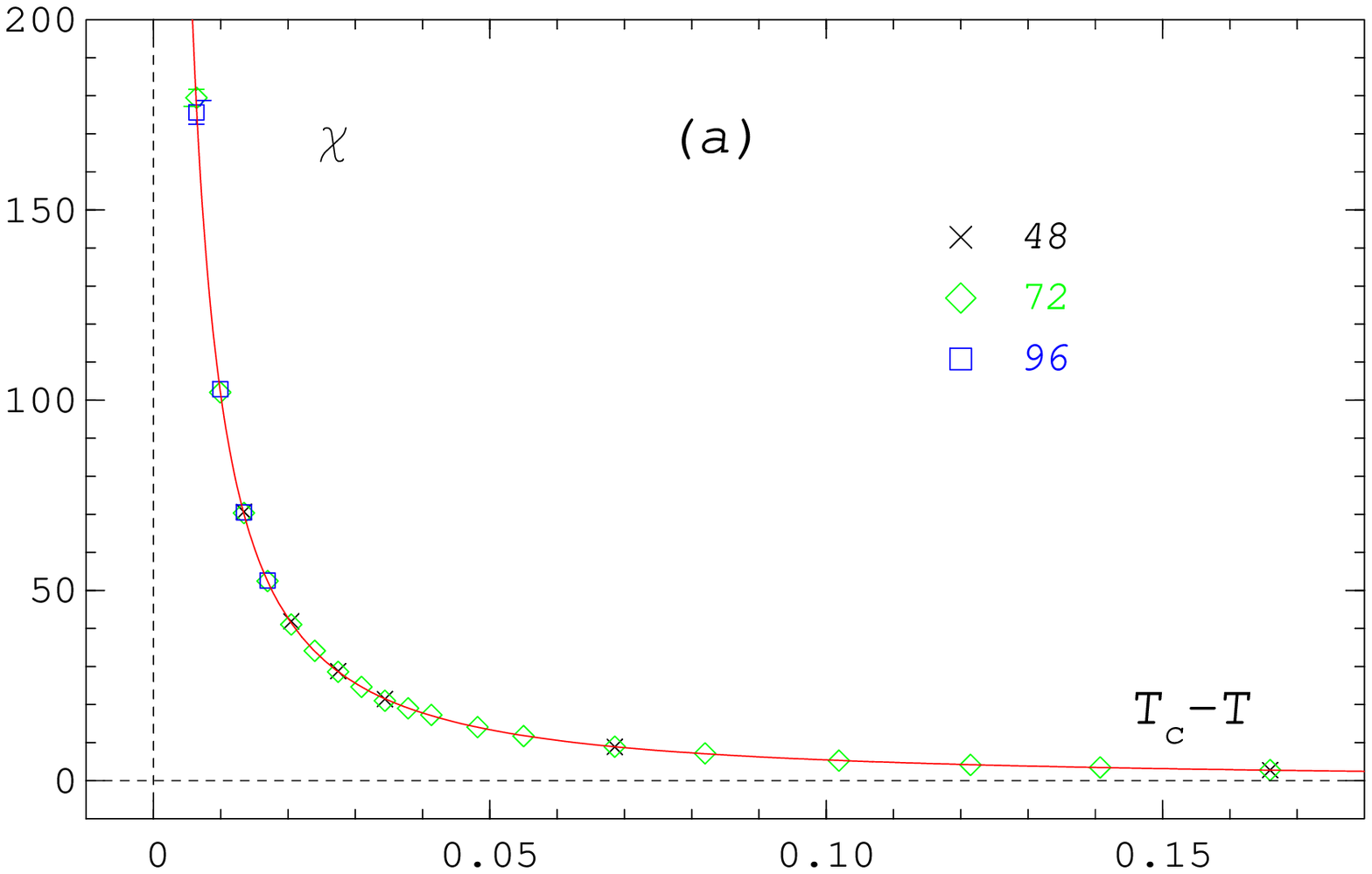, width=120mm}
\end{center}
\end{figure}
\begin{figure}[ht]
\begin{center}
   \epsfig{bbllx=63,bblly=280,bburx=516,bbury=563,
       file=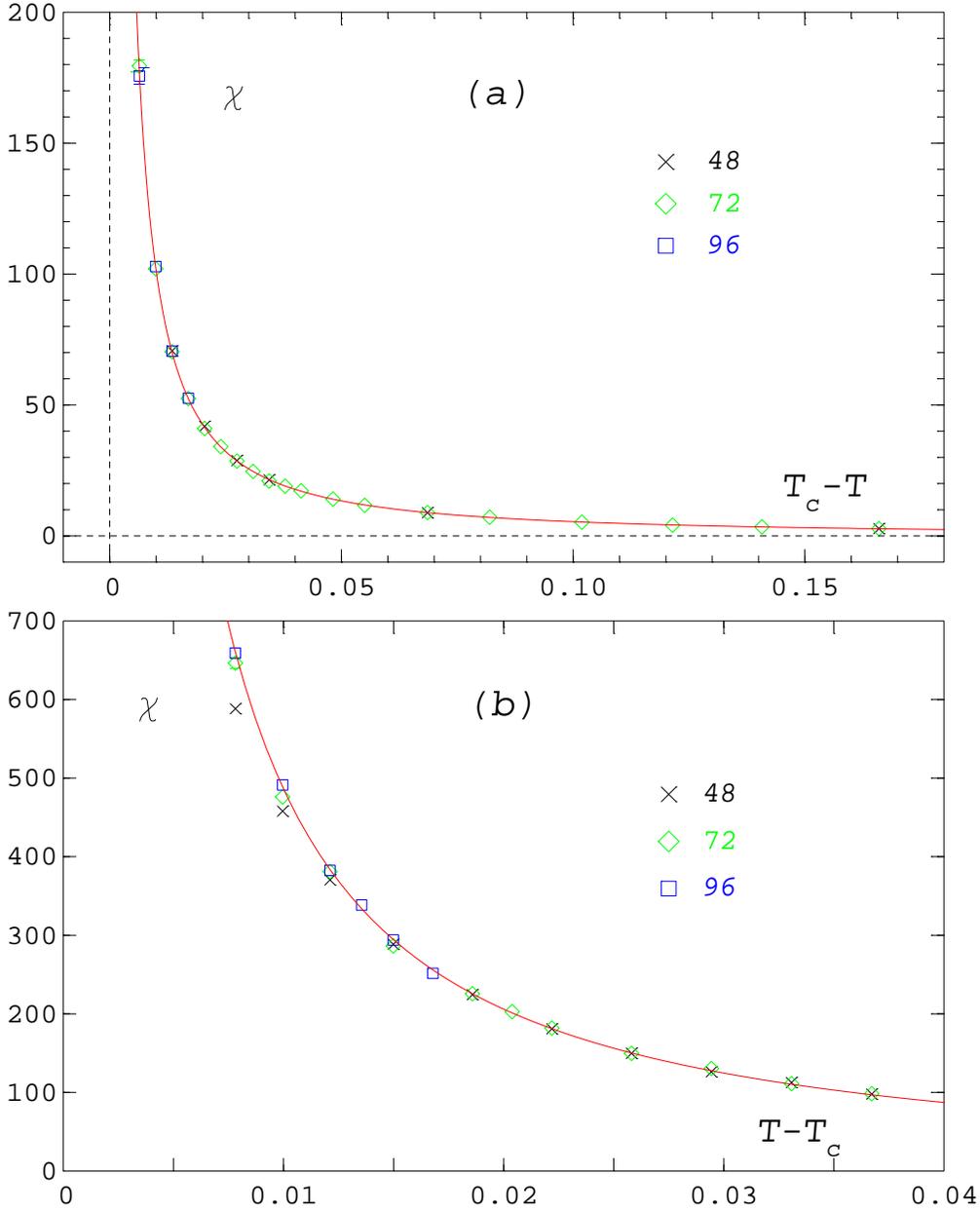, width=120mm}
\end{center}
\caption{The susceptibility $\chi$ at $H=0$ for $L=48,72$ and 96,
(a) below $T_c$ versus $T_c-T$ and (b) above $T_c$ versus $T-T_c$.
The solid lines show the fits from Eqs.\ (\ref{bchim}) and (\ref{bchip}),
respectively.}
\label{fig:chihzero}
\end{figure}
\n The remaining amplitudes $B_{1,2}^c$ may then be determined from a 
subsequent fit to all available data. As on the coexistence line we find
that the value for the leading non-analytic correction amplitude $B_1^c$
is small and of the same size as its error. We have therefore made fits
with fixed $B_1^c\equiv 0$ in the region  $0.00005 \le H \le 0.0025$ and find   
\be
M\; =\; 1.0435(15) H^{1/\delta} [1 - 2.65(5) H]~.
\label{bestmh}
\ee
with $\chi^2/N_f\approx 2\,$. From the value of $B^c$ in (\ref{bece}) we can
immediately deduce the values of two further quantities
\be
C^c\; =\;0.2172(3) \quad {\rm and} \quad D_c\: =\: H_0\: =\: 0.8150(56)~. 
\label{cede}
\ee
\indent Our next aim is the determination of the critical amplitudes $C^{\pm}$.
To this end we have evaluated the data of the susceptibility below and 
above the critical temperature for $H=0$. The data points are plotted
in Figs.\ \ref{fig:chihzero} (a) and (b) as a function of $(T_c-T)$ and
$(T-T_c)$, respectively. Inspired by our experience with the magnetization
in the cold phase, we have directly used a fit ansatz without a non-analytic
correction term. The data points from the largest lattices in the range
$0.01 \le T_c-T \:\lsim \: 0.10$ lead to 
\be
\chi\; =\; 0.1006(5) (-t)^{-\gamma} [1 -2.0(3)(-t)]~,
\label{bchim}
\ee
with $\chi^2/N_f\approx 1.3\,$. Note that here we have used the reduced
temperature as variable. A corresponding ansatz in the symmetric phase 
($T>T_c$) shows that the correction term is zero inside the error bars.
The points from the largest lattices are however well described by the 
leading term only with
\be
\chi\; =\; 0.4785(15)\: t^{-\gamma} ~,
\label{bchip}
\ee
Here, $\chi^2/N_f\approx 1.2\,$. From the fits (\ref{bchim}) and
(\ref{bchip}) we have
\be
C^- \: = \: 0.1006(5) \quad {\rm and} \quad C^+ \: = \: 0.4785(15)~.
\label{cepm}
\ee
This enables us to calculate the two universal amplitude ratios
\be
U_2\: = \: 4.756(28) \quad {\rm and} \quad R_{\chi}\: = \: 1.723(13) ~.
\label{fampr}
\ee
The ratio $U_2$ is in excellent agreement with two other MC results:
4.75(3) from the standard Ising model \cite{Caselle:1997hs} and 4.72(11)
from (3+1)-dimensional $SU(2)$ gauge theory \cite{Engels:1999nv}, as well
as other results (see Table 11 in \cite{Pelissetto:2000ek} and Table 5 in
\cite{Zinn-Justin:2001bf}). The ratio is also well inside the bounds
[4.688,4.891] derived in Ref.\ \cite{Bagnuls:2001jz}. The result
for $R_{\chi}$ is the first one solely from MC calculations. It is in
agreement with 1.57(23) from Refs.\ \cite{Zinn:1996sy,Fisher:fs} and
slightly higher than the result 1.660(4) from Refs.\
\cite{Campostrini:1999at,Campostrini:2002cf}.

\section{The Magnetic Equation of State}
\label{section:equofstate}

\begin{figure}[t!]
\begin{center}
   \epsfig{bbllx=127,bblly=265,bburx=451,bbury=588,
       file=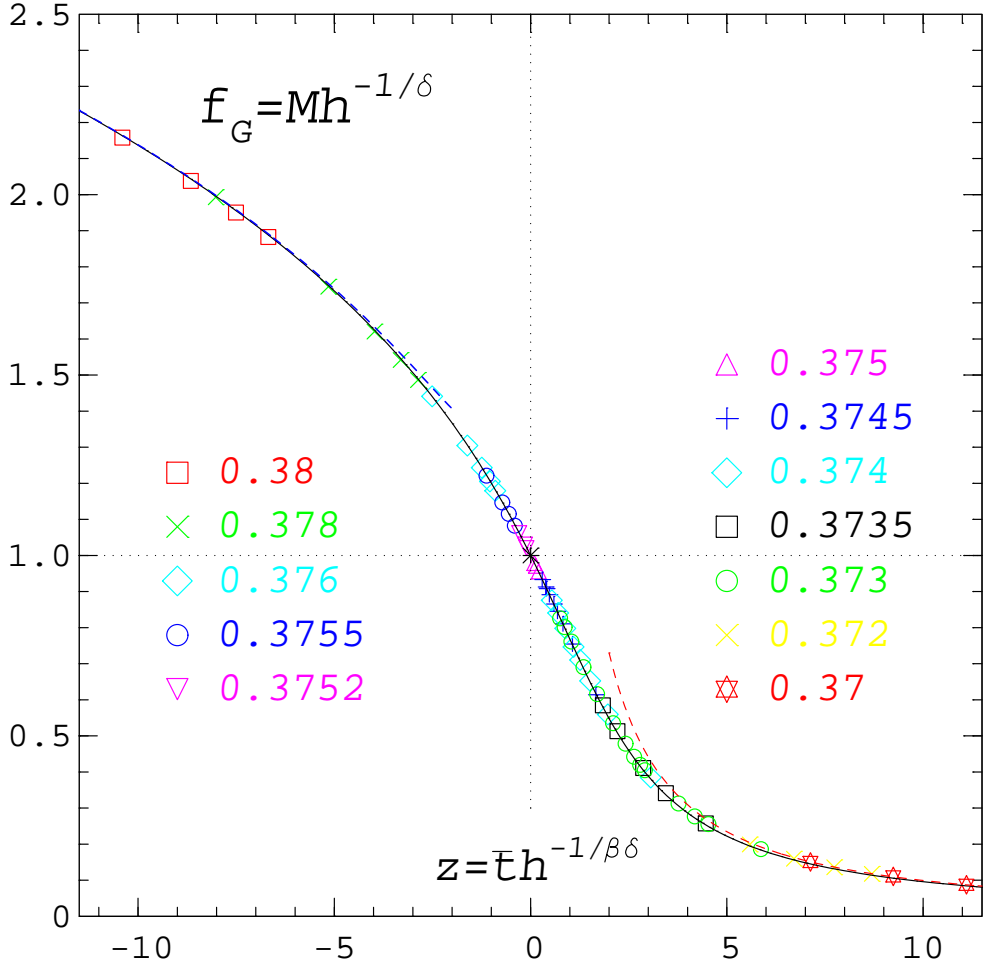, width=120mm}
\end{center}
\caption{The equation of state: $f_G=Mh^{-1/\delta}$ as a function of 
$z=\bar t h^{-1/\beta\delta}$. The solid line shows the parametrization
of Ref.\ \cite{Zinn-Justin:2001bf}, the dashed lines the asymptotic
forms (\ref{fgasym}) and (\ref{fgasyp}). The numbers refer to the 
different $J=1/T$-values of the data, the dotted lines cross at the
normalization point $f_G(0)=1$.}
\label{fig:fG(z)}
\end{figure}

\n After having determined the normalizations $T_0$ and $H_0$ we can 
calculate the universal scaling function of the magnetization for the
Ising class. To this end we have performed further simulations at 14 
couplings in the range $0.385\le J \le 0.365$ with external fields $H$
varying between 0.0001 and $0.005\,$. We convinced ourselves that the 
used lattice extensions were large enough to produce volume independent
results. For that purpose $L=48$ was sufficient at high temperatures 
(small $J$), closer to $T_c$ lattices with linear extension $L=72$ and 96 
were needed. The finite size effects diminished with increasing field,
generally they were stronger in the cold phase than in the hot phase. In
Fig.\ \ref{fig:fG(z)} we present the results for $f_G(z)$ obtained from
the magnetization data in the coupling range  $0.38\le J \le 0.37$.
Obviously these data are scaling well. Since the data from the couplings 
$J=0.385$ and 0.365 showed already slight deviations from scaling 
behaviour we have discarded them. Similarly, limitations were found for
the scaling $H$-regions. The data for $f_G(z)$ were scaling for $J=0.38$
only up to $H=0.0003$, for $J=0.378$ to $J=0.375$ up to $H=0.0005$ and 
for $J=0.3745$ to $J=0.372$ up to $H=0.0015$. At $J=0.37$ the scaling 
$H$-range was limited again by $H=0.0003$. Only data within these
parameter ranges are shown in Fig.\ \ref{fig:fG(z)}$\,.$ From the amplitude
ratios $U_2$ and $R_{\chi}$ in (\ref{fampr}) one can calculate the
asymptotic forms (\ref{fgasym}) and (\ref{fgasyp}) of $f_G(z)$. As can
be seen in Fig.\ \ref{fig:fG(z)}, the data for $f_G$ coincide with
the corresponding asymptotic form already for $z\:\lsim\: -4$ and
$z\:\gsim \:6-7$.

 The numerically determined scaling function can now be compared to 
parametric representations of the equation of state. A prominent and
widely accepted representation of the equation of state is that of Guida
and Zinn-Justin \cite{Zinn-Justin:2001bf}-\cite{Guida:1996ep},
which is based on field theory methods. They parametrize the equation of
state in terms of two new variables $R$ and $\theta$ by setting
\cite{Schofield:1969a}-\cite{Josephson:1969}
\ba
M\!\! &=&\!\! m_0 R^{\beta}\theta~, \label{linm}\\ 
\bar t\!\! &=&\!\! R(1-\theta^2)~, \label{lint}\\
h\!\! &=&\!\! h_0 R^{\beta\delta} \hat h(\theta)~, \label{linh}
\ea
where $m_0$ and $h_0$ are two normalization constants. Because $M$ is 
linear in $\theta$ this is also known as the linear model. The function
$\hat h(\theta)$ is an odd function and was approximated by a 5th order
polynomial with the following coefficients \cite{Zinn-Justin:2001bf} 
\be
\hat h(\theta) = \theta -0.76201(36)\theta^3 + 0.00804(11)\theta^5~. 
\label{hhat}
\ee
In Refs.\ \cite{Guida:1998bx,Guida:1996ep} slightly different
coefficients are given.
The function $\hat h(\theta)$ in (\ref{hhat}) has two zeros. The one at
$\theta=0$ corresponds to $h=0,\, t>0$, the second at $\theta_0=1.154$ 
to the coexistence line $h=0,\, t<0$. For $\theta=1$ one obtains the
critical line $t=0$. Using the normalization conditions (\ref{normf})
one finds
\be
m_0 = {(\theta_0^2-1)^{\beta} \over \theta_0}~, \quad
h_0 = {m_0^{\delta} \over \hat h (1)}~.
\label{m0h0}
\ee  
The Widom-Griffiths scaling variables $x$ and $y$ are only functions 
of $\theta$
\be
x = {1-\theta^2 \over \theta_0^2-1 } \left( {\theta_0 \over \theta}\right)
^{1/\beta}~,
\quad y = {\hat h (\theta) \over \hat h (1) \theta^{\delta} }~. 
\label{xandyth}
\ee
The same applies to the scaling function $f_G$ and its argument $z$
\be
z =  {1-\theta^2 \over \theta_0^2-1 }\theta_0^{1/\beta}
\left({\hat h (\theta) \over \hat h (1)}\right)^{-1/\beta\delta}~, \quad
f_G = \theta \left({\hat h (\theta) \over \hat h (1)}\right)^{-1/\delta}~.
\label{fgandzth}
\ee
In Fig.\ \ref{fig:fG(z)} the solid line represents the parametrization
from Eq.\ (\ref{fgandzth}) with $\hat h (\theta)$ from (\ref{hhat}).
Evidently, the agreement between the data and the solid line is 
excellent. Also, the asymptotic form for $z<0$ coincides directly with
the parametrization in a large range, whereas the asymptotic form for 
$z>0$ is marginally higher in the shown large $z$-range. 

\begin{figure}[ht]
\begin{center}
   \epsfig{bbllx=63,bblly=280,bburx=516,bbury=563,
       file=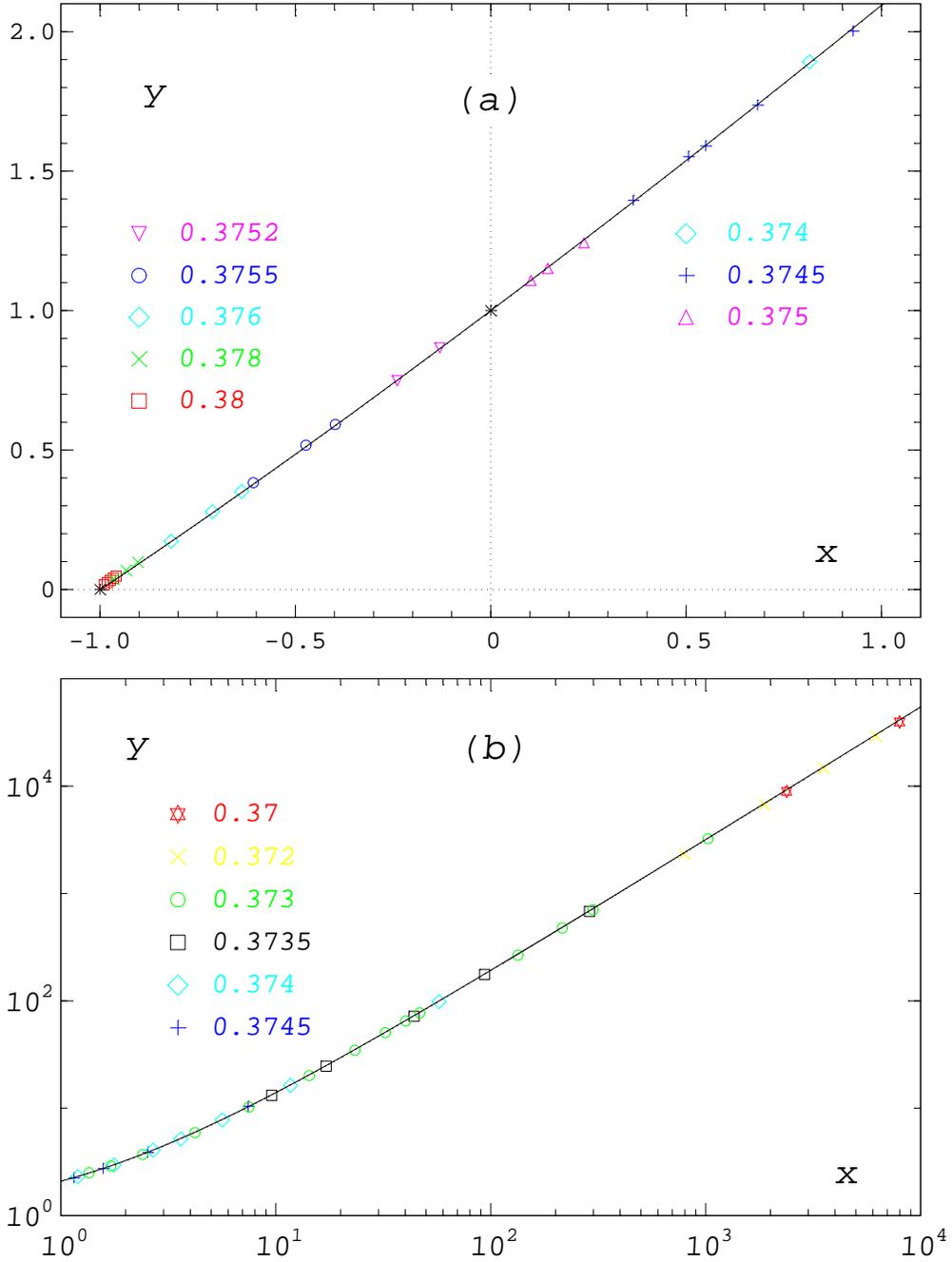, width=120mm}
\end{center}
\caption{The Widom-Griffiths form of the equation of state: $y=h/M^{\delta}$
as a function of $x=\bar t/M^{1/\beta}$. The solid lines show the
parametrization of Ref.\ \cite{Zinn-Justin:2001bf}, the numbers refer to 
the $J=1/T$-values of the data. Part (a) of the figure displays the small
$x$-region where the stars denote the normalizations (\ref{normf}),
part (b) the large $x$-region.}
\label{fig:y(x)}
\end{figure}

We have also plotted the data in the Widom-Griffiths form. The small
$x$-region is shown in Fig.\ \ref{fig:y(x)} (a), the large $x$-region in
Fig.\ \ref{fig:y(x)} (b). This kind of representation of the equation of
state is asymmetric with respect to the critical point: the cold or
broken phase ($T<T_c$) extends in $x$ from $-1$ to 0 and in $y$ from
0 to 1, the hot phase ($T>T_c$) reaches in $x$ from 0 to infinity and
in $y$ from 1 to infinity. Since the magnetization appears here in both 
\begin{figure}[t!]
\begin{center}
   \epsfig{bbllx=63,bblly=265,bburx=516,bbury=588,
       file=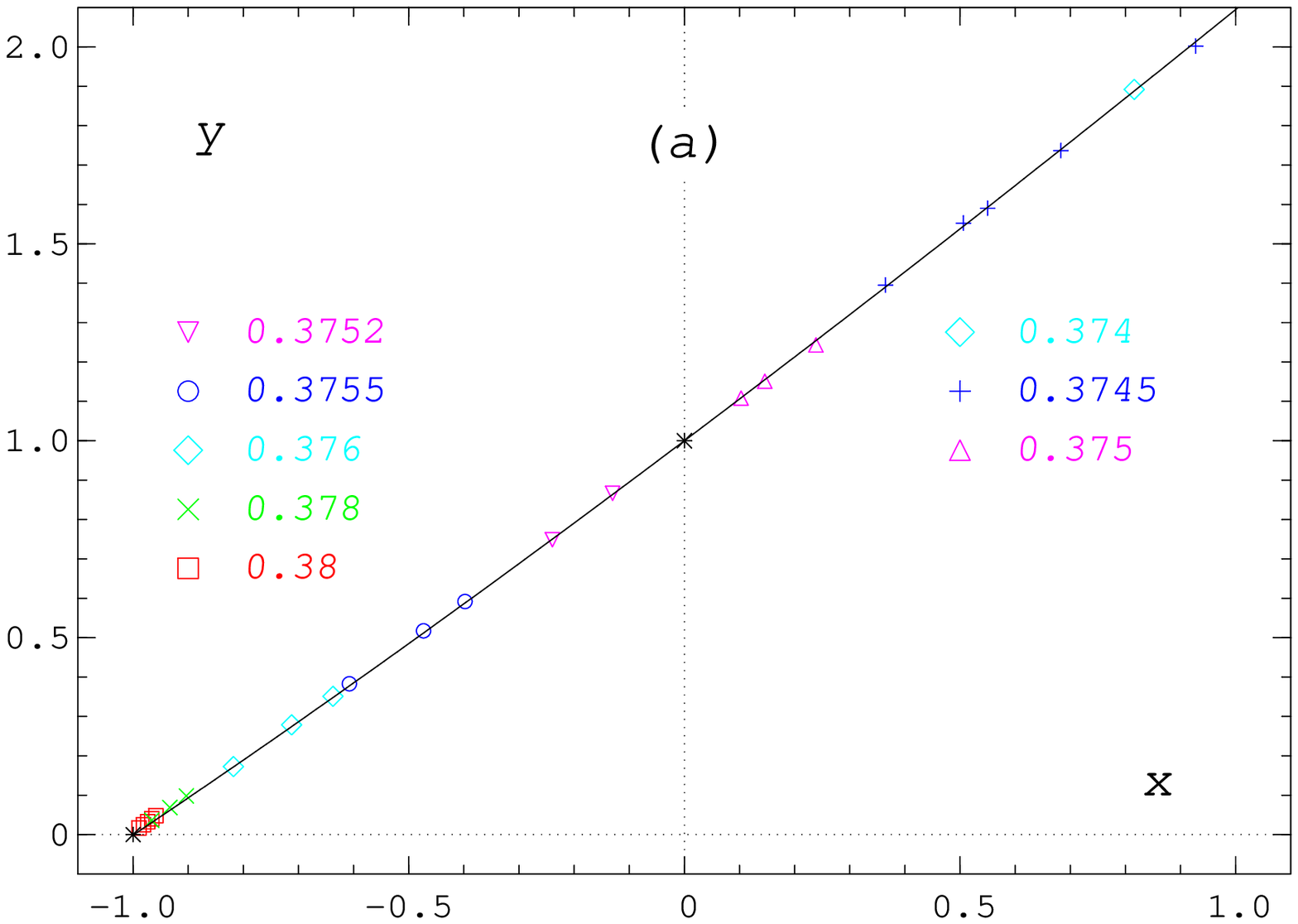, width=120mm}
\end{center}
\end{figure}
variables small deviations from scaling are easier to detect than in
Fig.\ \ref{fig:fG(z)}. This applies in particular to the cold phase. 
Therefore we included only points for $x<0$ in Fig.\ \ref{fig:y(x)} (a)
with $H \le 0.0003$, because the points with larger $H$-values
exhibited already very slight, but perceptible deviations. In the hot
phase we could use all the points already displayed in
Fig.\ \ref{fig:fG(z)}. As in Fig.\ \ref{fig:fG(z)}, the data show in
the Widom-Griffiths representation a similarly impressive agreement with 
the parametrization of the equation of state by Guida and 
Zinn-Justin.
\begin{figure}[t!]
\begin{center}
   \epsfig{bbllx=63,bblly=265,bburx=516,bbury=588,
       file=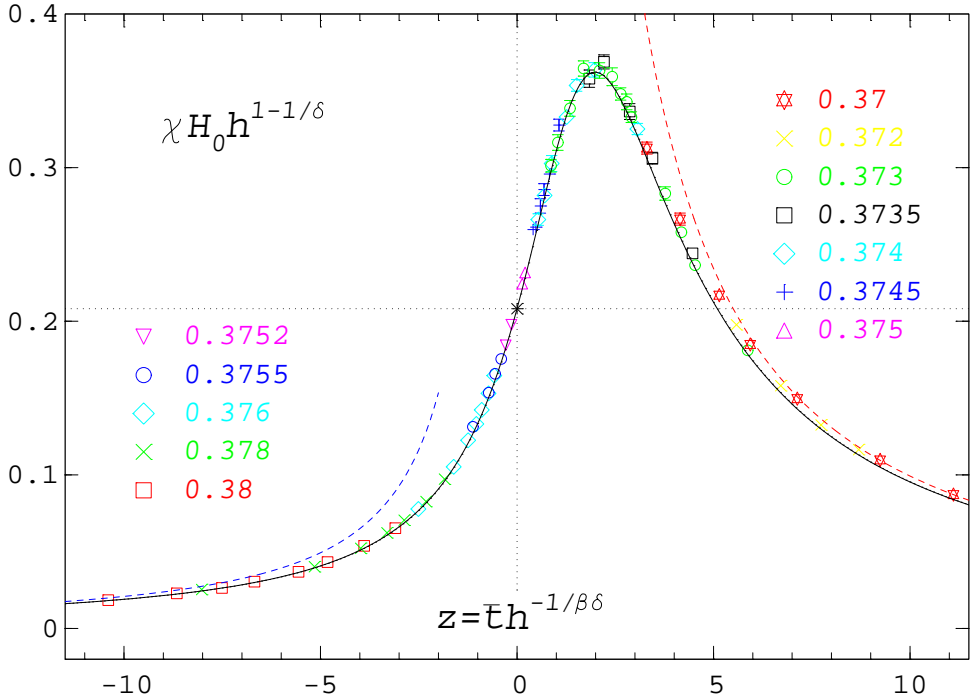, width=120mm}
\end{center}
\caption{The scaling function of the susceptibility 
$f_{\chi}(z)=\chi H_0 h^{1-1/\delta}$. 
The solid line is the parametrization of Ref.\ \cite{Zinn-Justin:2001bf}, 
the dashed lines are the asymptotic forms (\ref{fcasy}). The numbers refer 
to the $J=1/T$-values of the data. The star represents the normalization 
$f_{\chi}(0)=1/\delta$.}
\label{fig:fchi}
\end{figure}

 The scaling function $f_{\chi}(z)$ of the susceptibility is connected via
Eq.\ (\ref{fchi}) to the scaling function $f_G(z)$ of the magnetization. It
is therefore in principal known. On the other hand we have direct data for
$\chi$ from our simulations which allows for another check of parametric 
representations and the scaling hypothesis. In Fig.\ \ref{fig:fchi} we show
the respective data for the same $J$-values as in Fig.\ \ref{fig:fG(z)}.
We observe explicit scaling of the data, sometimes even in a larger $H$-range. 
In comparison to the parametrization of 
Ref.\ \cite{Zinn-Justin:2001bf} we find complete agreement in the cold 
phase ($z<0$) and a slightly increasing small difference to the data in the
hot phase ($z>0$). The reason for this little discrepancy lies in the
different asymptotic amplitudes which are used in the parametrization and
which we measured from our data. The amplitude in question is $R_{\chi}$.
In the parametrization scheme one derives
\be
R_{\chi} = \hat h(1) \theta_0^{\gamma/\beta} (\theta_0^2 -1)^{-\gamma}~.
\label{rchip}
\ee
This results in the number $R_{\chi}=1.666\,$, whereas we found from our
amplitude measurements on the critical line and at $h=0$ the value 
$R_{\chi}=1.723(13)$, that is a slightly higher number. The data shown in
Fig.\ \ref{fig:fchi} are consistent with that finding: at larger $z$-values
they coincide with the corresponding asymptotic curve.

\section{The Correlation Length}

\begin{figure}[t]
\begin{center}
   \epsfig{bbllx=63,bblly=280,bburx=516,bbury=563,
       file=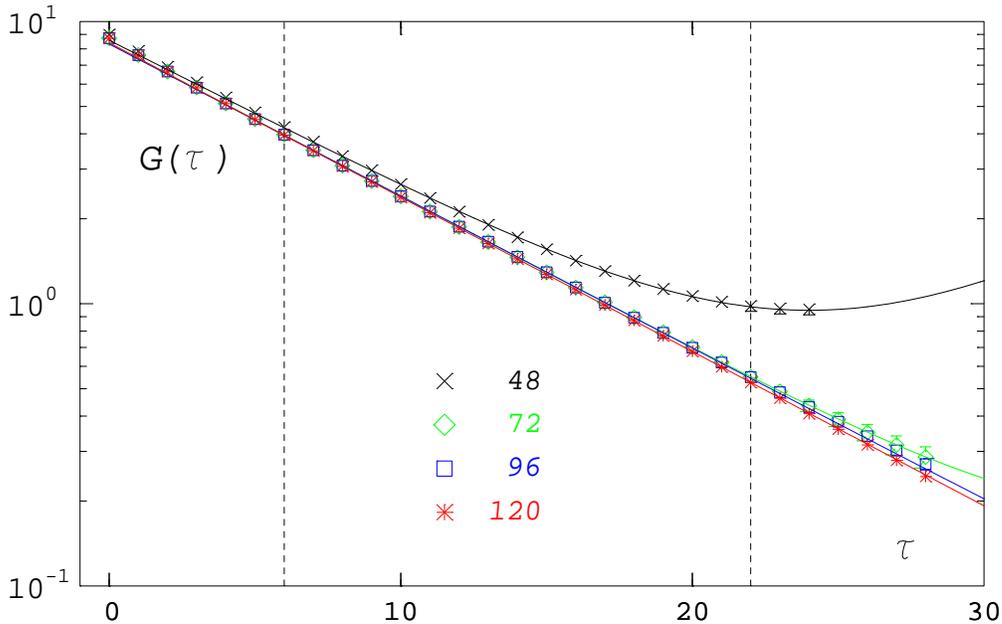, width=120mm}
\end{center}
\caption{The plane-correlation function $G(\tau)$ at $T=T_c$ and $H=0.0003$ 
for $L=48,72,96$ and 120. The solid lines show the fits from the ansatz
(\ref{expans}), the dashed lines indicate the $\tau$-range, inside which
the fits were done.}
\label{fig:G(tau)}
\end{figure}

Instead of using correlation functions of the individual spins
it is more favourable \cite{Engels:1994ek} to consider spin averages over 
planes and their respective correlation functions. For example, the spin
average over the $(x,y)$-plane at position $z$ is defined by
\be
S_z =  { 1 \over L^2 } \sum_{x,y} \phi_{x,y,z}~.
\label{spinav}
\ee
The average $S$ of all $S_z$ is equal to the lattice average $\phi$
and 
\be
\langle S_z \rangle =\langle S \rangle =\langle \phi \rangle~. 
\label{sexp}
\ee 
Correspondingly, we define the plane-correlation function $G(z)$ by
\be
G(z) = L^2 (\langle S_0 S_z \rangle - \langle S \rangle^2)~.
\label{planecf}
\ee
Here, $z$ is the distance between the two planes. Instead of choosing
the $z$-direction as normal to the plane one can as well take the 
$x$- or $y$-directions. Accordingly, we enhance the accuracy of the 
correlation function data by averaging over all three directions 
and all possible translations. The correlators are symmetric
and periodic functions of the distance $\tau$ between the planes
\be
G(\tau) = G(-\tau)~, \quad {\rm and}\quad  G(\tau) = G(L-\tau)~.
\label{symper}
\ee
The factor $L^2$ on the right-hand side of (\ref{planecf}) ensures
the relation
\be
\chi = \sum_{\tau=0}^{L-1} G(\tau)~ = \sum_{\tau=-L/2+1}^{L/2} G(\tau)~.
\label{fluc}
\ee
Similarly, the second moment $\mu_2$ may be expressed in terms of the 
plane correlators as
\be
\mu_2 = d \sum_{\tau=-L/2+1}^{L/2} \tau^2 G(\tau)~.
\label{secm}
\ee
Like the point-correlation functions in Eq.\ (\ref{xiexp}) the 
plane correlators $G(\tau)$
decay exponentially. In Fig.\ \ref{fig:G(tau)} we show the logarithm of 
typical plane-correlation functions for different $L$, here at $T=T_c$
and $H=0.0003$. The expected behaviour is actually confirmed, as well as
the influence of periodic boundary conditions on finite lattices. 
In order to obtain the exponential correlation length from $G(\tau)$ 
on a lattice with linear extension $L$ and periodic boundary conditions
we therefore make the ansatz
\be
G(\tau ) = A \left[\: \exp (-\tau /\xi ) + 
\exp (-(L-\tau )/\xi )\: \right]~,
\label{expans}
\ee
and then try to fit the data for $G(\tau)$ in an appropriate $\tau$-range.
The ansatz (\ref{expans}) implies of course, that there are no additional 
excitations contributing to $G(\tau)$. Inspired by the experiences reported
in \cite{Caselle:1997hs}, we proceed in the following way. First we 
calculate an effective correlation length $\xi_{eff}(\tau)$ from 
(\ref{expans}) using only the correlators at $\tau$ and $\tau +1$. For
$\tau \ll L$ this correlation length is approximately given by
\be
\xi_{eff}(\tau) = {-1 \over \ln (G(\tau +1)/G(\tau)) }~.
\label{xieff}
\ee
\n With increasing $\tau$ also $\xi_{eff}$ increases and eventually reaches 
a plateau inside its error bars. Most likely, the lower values of 
$\xi_{eff}$ at small $\tau$ are due to higher excitations. At large 
distances $\tau$ the resulting $\xi_{eff}$ start to fluctuate when the
relative error of the data becomes too large. The two limits define an
intermediate $\tau$-range where a global fit with the ansatz (\ref{expans})  
can finally be used to estimate the exponential correlation length.

 In order to give an impression of the actual implementation of this
procedure we show in Table \ref{tab:xi} the results of the fits 
($\xi^{fit}$) to the plane-correlation functions from 
Fig.\ \ref{fig:G(tau)}. The corresponding correlation length is here
close to 8. The plateau of $\xi_{eff}$ is reached for all $L$ at a value
for $\tau_{min}$ between 6 and 10. A reasonable upper fit limit is 
$\tau_{max} =18$ to 22.
Varying $\tau_{max}$ in this range does not affect the fit result.
Also a variation of $\tau_{min}$ from 6 to 10 leads only to changes
inside the error bars. The respective $\chi^2/N_f$ are less than or equal
to 0.02. If we compare $\xi^{fit}$ to the effective correlation length 
at $\tau_{min}$, we find essentially the same result, including the size 
of the error bar.

We have calculated as well the correlation length $\xi^F$ from Eq.\ 
(\ref{xif}). We see from Table \ref{tab:xi}, that $\xi^F$ is compatible
with $\xi^{fit}$, but has a larger error. Only for the largest lattice
with $L=120$ we obtain a difference. This may be due to the use of all
correlations in the calculation of the Fourier transform, also those
at large distances, where the relative errors are large. In addition,
the numerical determination of $\xi^F$ requires an increasing accuracy 
with increasing $L$ because of the diminishing denomiator $\sin(\pi/L)$
in  Eq.\ (\ref{xif}) and the difference $\chi/F-1$ in the nominator.

\begin{table}[t]
\begin{center}
\begin{tabular}{|r||r|c|l|c||c|c|}
\hline
$L$ & $\tau_{min}$ & $\tau_{max}$ & $\xi^{fit}$ &
 $\xi_{eff}(\tau_{min})$ & $\xi^F$ & $\xi_{2nd}$ \\
\hline
  48 &  7 & 18 & 8.265(104) & 8.278(79) & 8.18(16) &  6.80(11) \\
  72 &  6 & 20 & 7.986(45)  & 7.917(50) & 7.92(15) &  7.52(17) \\
  96 &  8 & 19 & 8.055(50)  & 8.011(49) & 8.18(16) &  8.10(21) \\
 120 & 10 & 22 & 7.957(44)  & 7.948(49) & 7.43(16) &  7.51(10) \\
\hline
\end{tabular}
\end{center}
\vspace{-0.2cm}
\caption{The correlation length at $T=T_c$ and $H=0.0003$ from different 
methods: $\xi^{fit}$ from fits to the ansatz (\ref{expans}) in the 
$\tau$-interval $[\tau_{min},\tau_{max}]$; $\xi^F$ from Eq.\ (\ref{xif}) 
and $\xi_{2nd}$ from Eq.\ (\ref{xi2nd}) using Eq.\ (\ref{secm}).
For comparison we show also $\xi_{eff}(\tau_{min})$.}
\label{tab:xi}
\end{table}

The accurate numerical calculation of $\xi_{2nd}$ is also problematic.
The reason for that is the weight $\tau^2$ in the sum of Eq.\ (\ref{secm}),
which emphasizes the correlators $G(\tau)$ with the largest relative errors.
Caselle and Hasenbusch \cite{Caselle:1997hs} have therefore replaced the 
correlation function for $\tau>\tau_{max}\approx 3\xi$ by its large 
distance form with $\xi=\xi_{eff}(\tau_{max})$. The sum was then extended
to infinity. The procedure provides an estimate of the infinite volume
value of $\mu_2$. On the other hand the finite size dependence of
$\xi_{2nd}$ is then lost and the determination of $\xi_{2nd}$ is no longer
independent of the other definitions. Because of that, we have derived the
second moment correlation length directly from Eq.\ (\ref{secm}),
truncating the sum when the numerical correlation function result was
compatibel with zero. The corresponding values of $\xi_{2nd}$ in 
Table \ref{tab:xi} have relatively large errors. The values increase with
increasing $L$, approaching the results from other definitions from below.
At $L=120$ the estimate for $\xi_{2nd}$ is probably too low, because of the
mentioned numerical difficulties. 

Our test of the different definitions of $\xi$ was performed at $T=T_c\,$.
Already here we have found indications for the presence of higher states. 
In general, the influence of higher exitations becomes even stronger in 
the broken phase. An extensive discussion of this point and the actual
determination of the first two higher states with a variational technique
can be found in Ref.\ \cite{Caselle:1999tm}. In the same paper and in
Ref.\ \cite{Caselle:1997hs} the numerical estimates of the exponential
correlation length in the broken phase have been compared to high precision
results from the $Z_2$ gauge model \cite{Agostini:1996xy}, which is related
to the Ising model by a duality transformation. The corresponding values
differed at most in the third digit and showed that $\xi_{eff}$ slightly
underestimates the exponential correlation length. In the high temperature 
phase no nearby exitations are present and $\xi_{eff}$ is a good estimator
of the true correlation length. As a result of these considerations we 
resort in the following to the estimation of the exponential correlation
length via the fit method. The corresponding scaling function can be
calculated to a good approximation in this way. The extreme precision 
aimed at in Refs.\ \cite{Caselle:1997hs} and \cite{Caselle:1999tm} is not
required for that purpose. Moreover, a possible small underestimation of $\xi$
in the broken phase can be controlled by comparing the critical amplitude
ratio $\xi^+/\xi^-$, which we calculate below, to the known results for 
the ratio $U_{\xi_{exp}}$.

\subsection{The Critical Amplitudes of the Correlation Length}
\label{subsection:conclusion}

\begin{figure}[t!]
\begin{center}
   \epsfig{bbllx=63,bblly=280,bburx=516,bbury=563,
       file=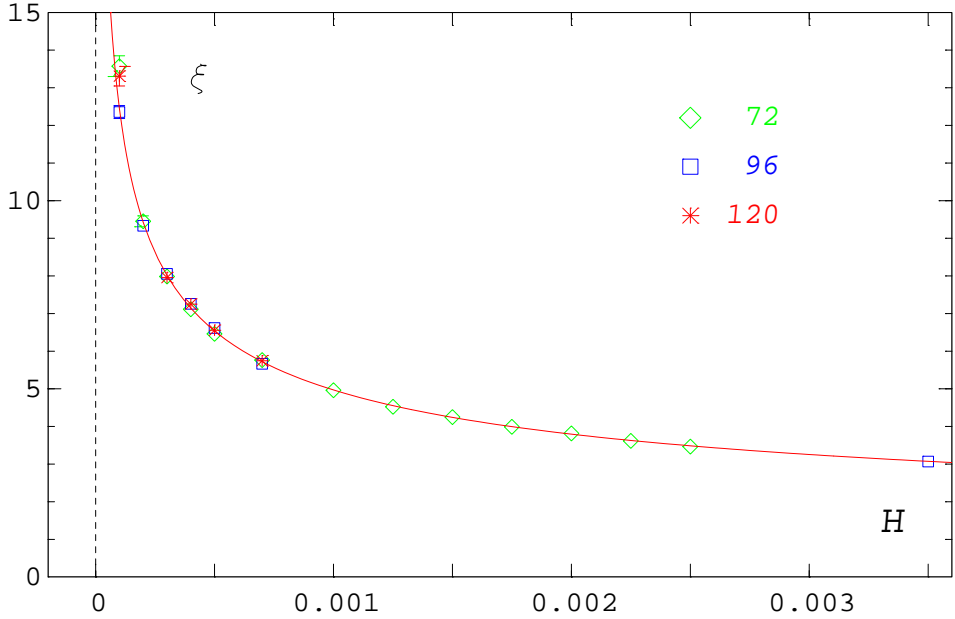, width=120mm}
\end{center}
\caption{The correlation length $\xi$ at $T_c$ as a function of the magnetic
field $H$ for $L=72,96$ and 120. The solid line shows the fit (\ref{xicfit}).}
\label{fig:xiah}
\end{figure}
\begin{figure}[ht]
\begin{center}
   \epsfig{bbllx=63,bblly=280,bburx=516,bbury=563,
       file=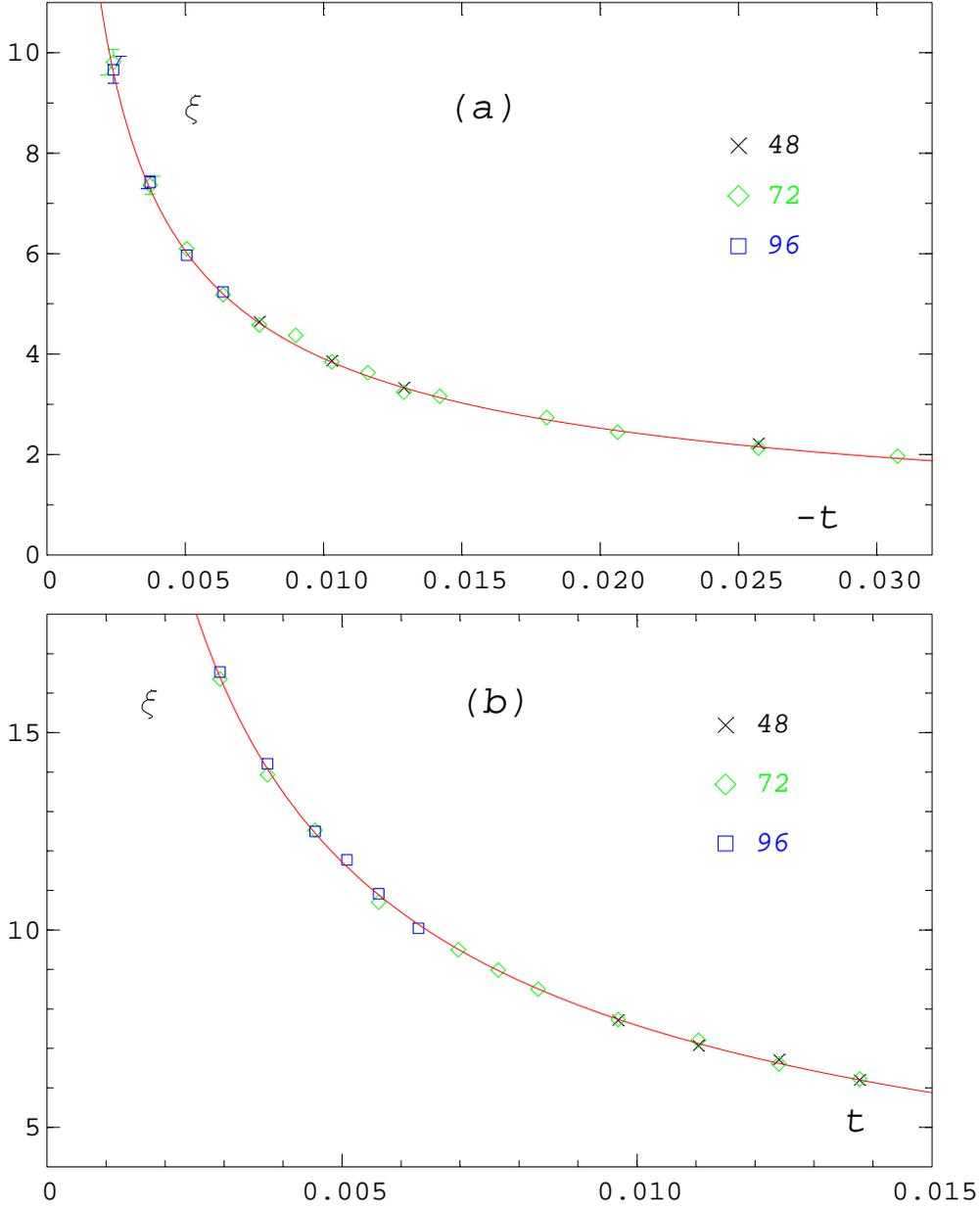, width=120mm}
\end{center}
\caption{The correlation length $\xi$ at $H=0$ for $L=48,72$ and 96,
(a) below $T_c$ versus $-t$ and (b) above $T_c$ versus $t$.
The solid lines show the fits from Eqs.\ (\ref{ximamp}) and (\ref{xipfit}),
respectively.}
\label{fig:xiat}
\end{figure}

\noindent We have determined the correlation length on the critical isotherm
for $H>0$ on lattices with $L=72,96$ and 120. The results are shown in Fig.\
\ref{fig:xiah}. We have performed fits to these data with the ansatz
\be
\xi \;=\; \xi^c H^{-\nu_c} [1+\xi^c_1 H^{\omega\nu_c} +\xi^c_2 H ]~.
\label{xianh}
\ee
As in the case of the magnetization the first correction-to-scaling term
turned out to be compatible with zero. Subsequently, we have therefore made
fits with $\xi^c_1\equiv 0$ in the interval $0.0002 \le H \le 0.0035$.
Irrespective of whether we use all data in the interval, or only those from 
the largest lattices or subsets of them we find
\be
\xi = 0.3048(9) H^{-\nu_c} [1 + 9.8(2.5) H ]~,
\label{xicfit}
\ee 
with $\chi^2/N_f$ varying between 0.8 and $1.1\,$. The corresponding critical
amplitude is then
\be
\xi^c = 0.3048(9)~.
\label{xicamp}
\ee
The correlation length data for $H=0$ below and above the critical temperature
are shown in Figs.\ \ref{fig:xiat} (a) and (b). They are slighly more
fluctuating than the data in Fig.\ \ref{fig:xiah}. Nevertheless, the 
corresponding critical amplitudes may be determined quite well. Here, we
have tested again that a fit ansatz without the first correction-to-scaling
term is sufficient
\be
\xi = \xi^{\pm} |t|^{-\nu} [ 1 + \xi^{\pm}_2 t]~.
\label{xiant}
\ee  
\begin{figure}[t!]
\begin{center}
   \epsfig{bbllx=63,bblly=280,bburx=516,bbury=563,
       file=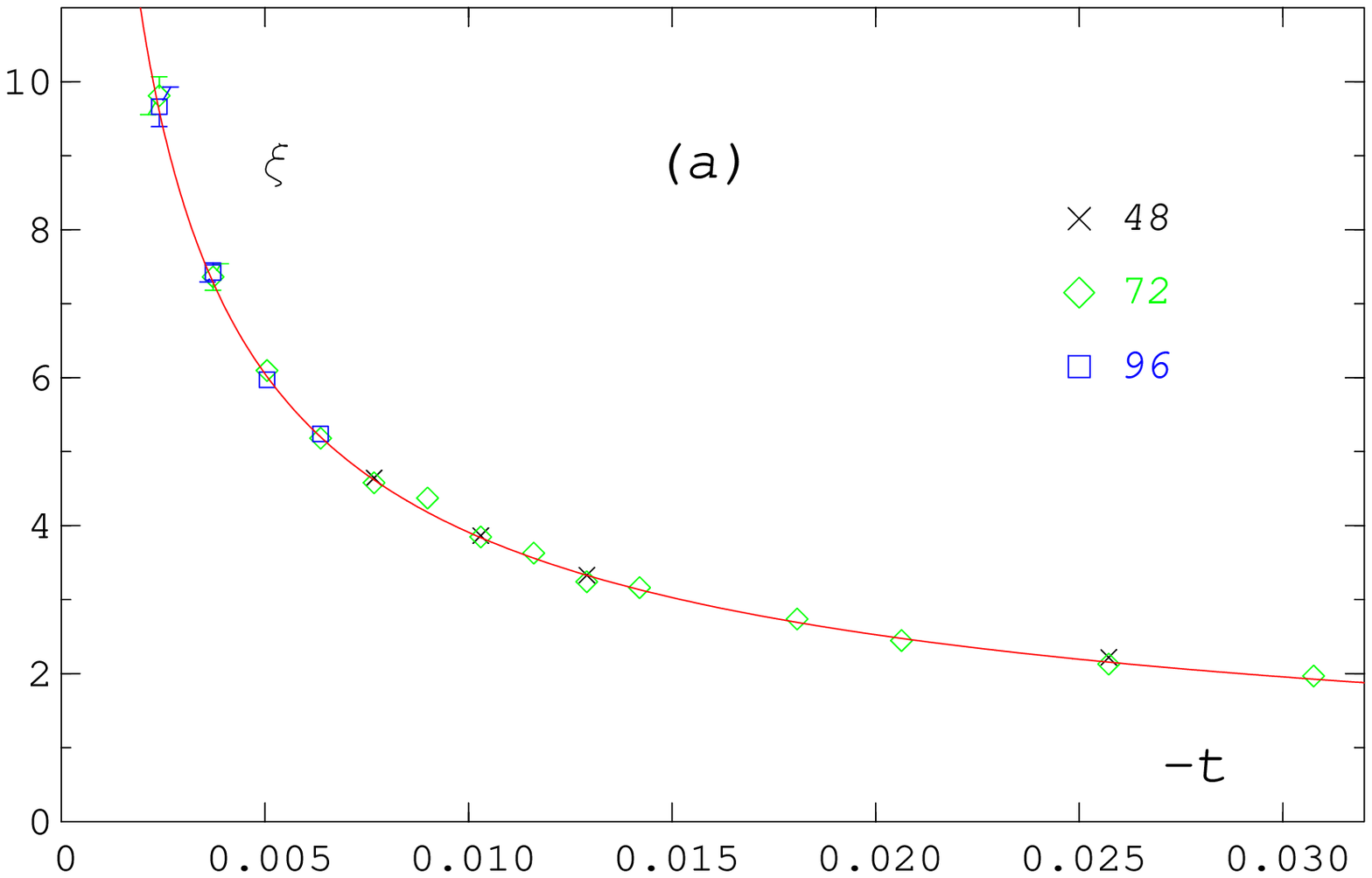, width=120mm}
\end{center}
\end{figure}
\noindent As it turns out, the best fits in the broken phase (with 
$\chi^2/N_f\approx 3$) are those using all points and setting 
$\xi^-_2 \equiv 0$. We obtain 
\be
\xi = 0.2145(15)\cdot (-t)^{-\nu}, \quad {\rm that~is}
\quad \xi^- = 0.2145(15)~.
\label{ximamp}
\ee
The error of $\xi^-$ includes also all other fit results, using only the
data from the largest lattices and/or with a correction term proportional to
$t$. In the symmetric phase the correction term is only observable, when
the larger $t$-points are included in the fits. We have therefore first 
performed fits in the interval $0.002 < t < 0.009$ omitting the correction
term and find for the critical amplitude the value
\be
\xi^+ = 0.415(1)~.
\label{xipamp}
\ee
The subsequent fit with all data leads to
\be
\xi = 0.415(1) t^{-\nu} [ 1 + 0.2(1) t]~,
\label{xipfit}
\ee
with again $\chi^2/N_f\approx 3$. From the critical amplitudes for $H=0$
we derive the ratio $\xi^+/\xi^- = 1.935(14)$. As discussed already, this 
number is to be compared to the results 1.891(22) from \cite{Caselle:1997hs},
\cite{Caselle:1999tm,Agostini:1996xy} and 1.896(10) from 
\cite{Campostrini:1999at,Campostrini:2002cf} for this ratio. 
We deduce from this check, that our $\xi$-values in the broken phase are
about $2\%$ lower than the true exponential correlation length. The critical
amplitudes $\xi^+$ and $\xi^c$, which we have also determined, should however
be unaffected by possible higher states \cite{Caselle:1999tm}. That allows us
to calculate the universal ratios $Q_c$ and $Q_2$ from Eqs.\ (\ref{rce}) and
(\ref{hratios}). We obtain
\be
Q_c = 0.326(3) \quad {\rm and} \quad Q_2 = 1.201(10)~.
\label{Qs}
\ee   
Our finding for $Q_c$ is well in accord with two other results: 0.328(5) from
\cite{Caselle:1997hs} and 0.324(6) from \cite{Fisher:fs}, it is slightly below
the value 0.3315(10) from \cite{Campostrini:1999at,Campostrini:2002cf}. The
number for $Q_2$ is the first one from MC simulations and compares well with
1.195(10) from \cite{Campostrini:1999at,Campostrini:2002cf}, the result 
1.17(2) from \cite{Zinn:1996sy,Fisher:fs} is somewhat lower.  
\section{The Scaling Function of the Correlation Length}
\label{section:scalefunc}
\begin{figure}[tp]
\begin{center}
   \epsfig{bbllx=63,bblly=265,bburx=516,bbury=588,
       file=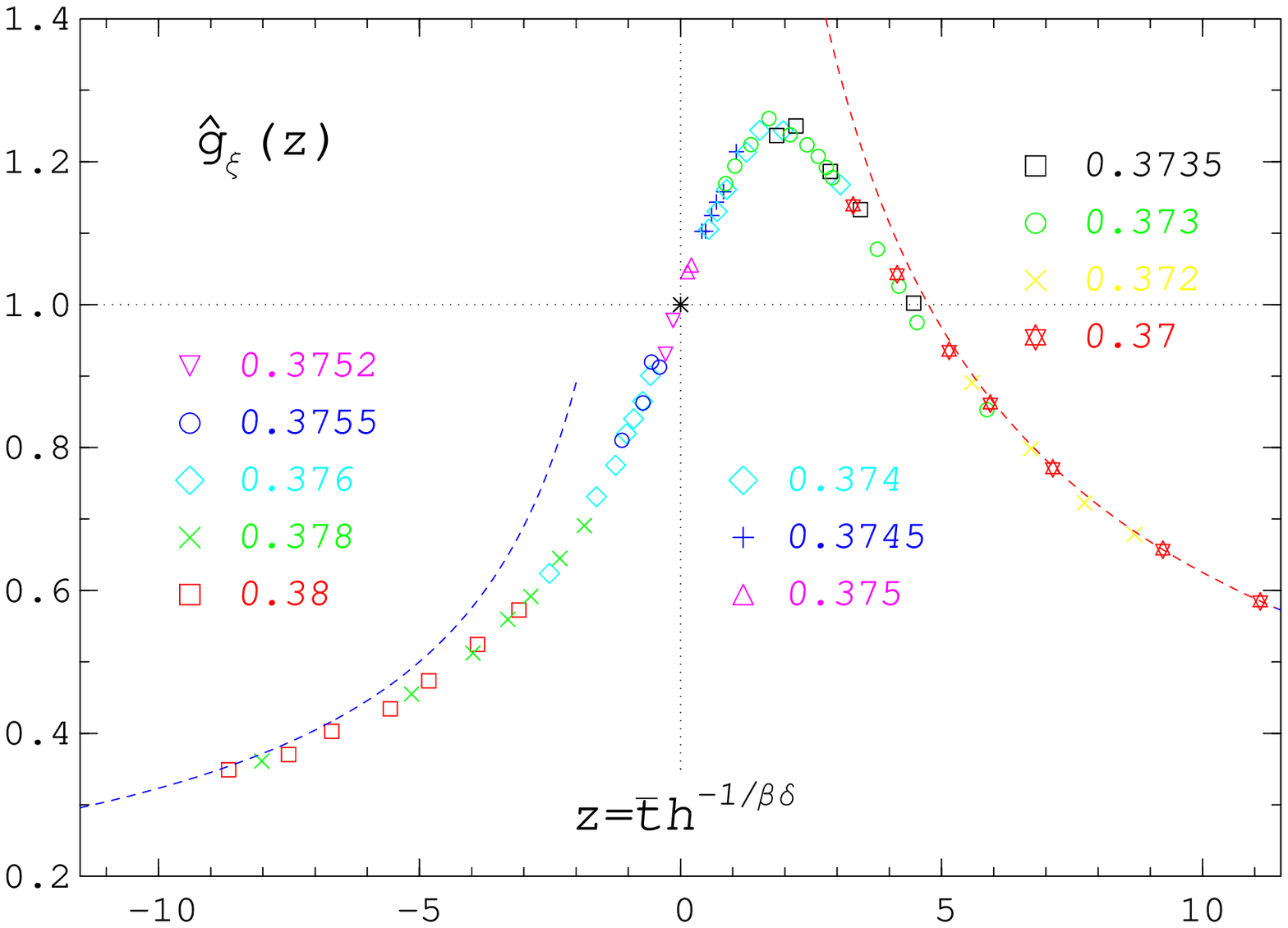, width=120mm}
\end{center}
\caption{The scaling function of the correlation length:
$\hat g_{\xi}(z)=\xi h^{\nu_c}/g_{\xi}(0)$. The dashed lines are the 
asymptotic forms calculated from Eq.\ (\ref{gxiasy}). The numbers refer
to the different $J=1/T$-values of the data.}
\label{fig:gexi}
\begin{center}
   \epsfig{bbllx=63,bblly=265,bburx=516,bbury=588,
       file=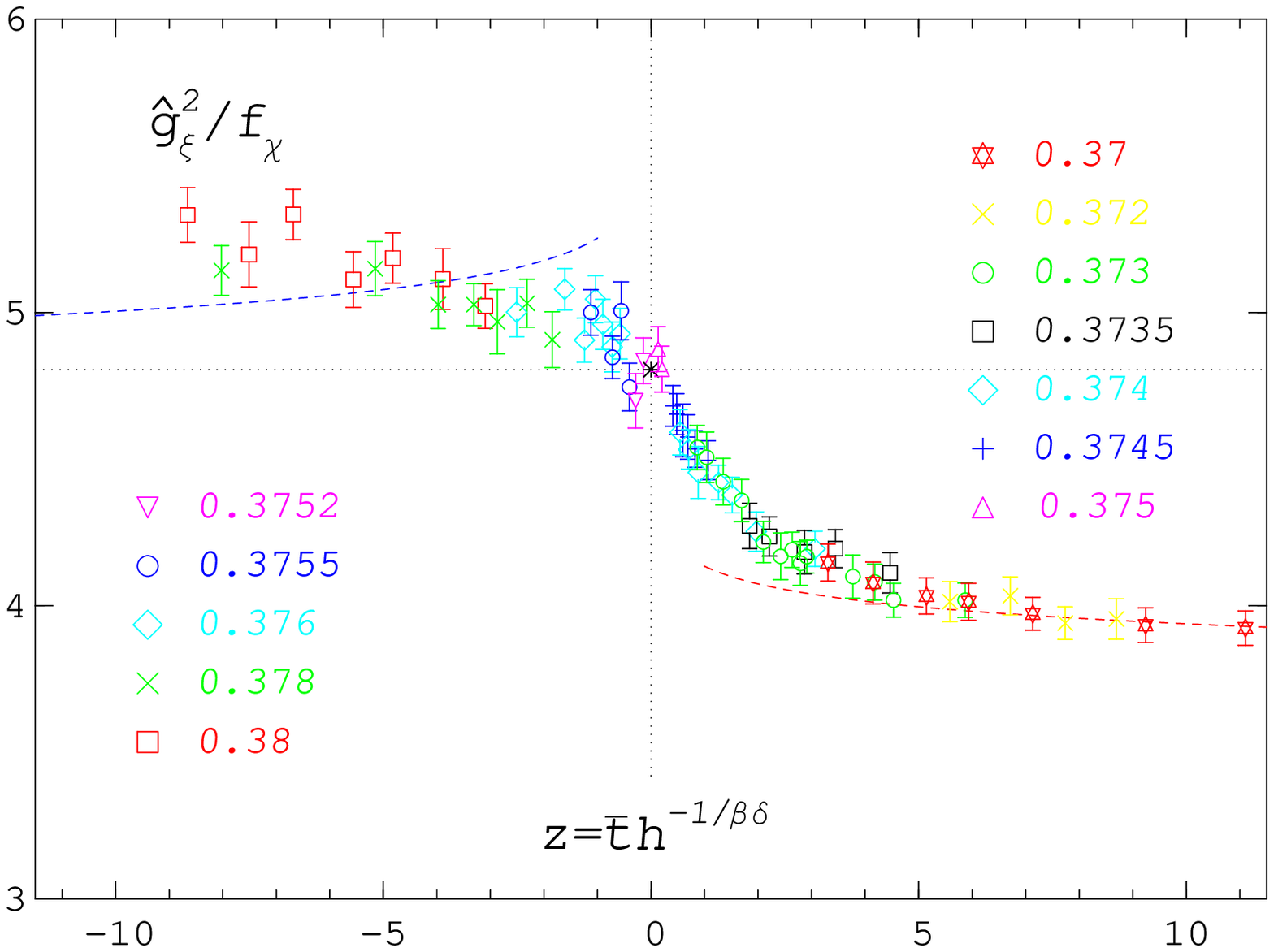, width=120mm}
\end{center}
\caption{The ratio $\hat g_{\xi}^2 /f_{\chi}(z)$ of scaling functions.
The dashed lines are the asymptotic forms calculated from Eqs.\
(\ref{ratasy}) and (\ref{ratamp}). The numbers refer to the 
different $J=1/T$-values of the data, the star shows the normalization
point $\hat g_{\xi}^2/f_{\chi}(0)=\delta$.}
\label{fig:rat}
\end{figure}

\noindent In Fig.\ \ref{fig:gexi} we show the data for the normalized scaling 
function $\hat g_{\xi}(z)=\xi h^{\nu_c}/g_{\xi}(0)$. The determination of 
the correlation length was sometimes difficult, in particular in the region
below $T_c$, because the correlation function data were not always accurate
enough to identify a reasonable plateau in $\xi_{eff}$. In most of these 
cases we could resort to the results of smaller lattices, after convincing
ourselves from data adjacent in $H$, that no finite size effects were present.   
The data shown in Fig.\ \ref{fig:gexi} correspond to the ones presented in 
Fig.\ \ref{fig:fchi} for the scaling function of the susceptibility, apart 
from a point at $z=-10.4\,$. The normalization was calculated from
Eq.\ (\ref{gxi0}) and is  $g_{\xi}(0)=0.331(1)$. As in the case of the
magnetization and the susceptibility we observe scaling of the correlation
length data in a large $z$-range and an early approach to the asymptotic
forms calculated from Eq.\ (\ref{gxiasy}). The shape of $\hat g_{\xi}(z)$
is very similar to that of $f_{\chi}(z)$ - both functions have a peak at
about the same $z$-value, slightly below 2. It is therefore natural to
consider the ratio $\hat g_{\xi}^2/f_{\chi}$ appearing in Eq.\ (\ref{rxichi}).
Taking the square of $\hat g_{\xi}$ instead of the function itself leads
to a very slowly varying function of $z$. That can be seen from the
asymptotic behaviour
\be
\hat g_{\xi}^2/f_{\chi} (z) \; {\raisebox{-1ex}{$\stackrel 
{\displaystyle =}{\scriptstyle z \rightarrow \pm\infty}$}} \;
r^{\pm} (\pm z)^{-\eta\nu}~.
\label{ratasy}
\ee
\noindent Here, $\eta\nu =0.0211$ and  
\be
r^+ = \left[ \left( {\delta \over Q_2} \right)^2 R_{\chi}^{\;\,\eta}
\right]^{1/( 2-\eta)} \quad 
{\rm and} \quad {r^+ \over r^-} = { U_{\xi}^2 \over U_2 }~.
\label{ratamp}
\ee
\noindent For $\;\eta \rightarrow 0$ one obtains an even simpler picture: the 
ratio $\hat g_{\xi}^2/f_{\chi}$ attains the constant values $r^{\pm}$ for 
$(\pm z) \rightarrow \infty$, where $r^+ = \delta /Q_2=4.001$ and $r^+/r^-$ 
is very close to $3/4$ (when one takes the "true" value $U_{\xi}=1.89$),
at $z=0$ one has $\hat g_{\xi}^2/f_{\chi}= \delta=4.805$. In Fig.\
\ref{fig:rat} we show the behaviour of the ratio from the actual data and 
we compare it to the asymptotic forms (\ref{ratasy}) derived from the
amplitudes which we have determined. As expected, the ratio is slowly
varying, apart from the close neighbourhood of the critical line $z=0$. In
the hot phase ($z>0$) we obviously have a consistent behaviour of the data
and a clear, early approach to the asymptotic curve, whereas in the cold 
phase ($z<0$) the data are more noisy, probably because of the difficulties
in calculating the correlation length with sufficient precision there. We refrain
here from a direct comparison with the parametric representation of Fisher
et al.\ \cite{Fisher:1999tr} for $\xi^2/2\chi$ because we use a slightly
different set of critical exponents (those of Ref.\ 
\cite{Zinn-Justin:2001bf}), which is crucial here. Moreover, since we
consider an improved instead of the standard Ising model the relative 
normalizations of the variables are unclear.

\section{Summary and Conclusions}
\label{section:conclusion}
In this paper we have calculated the equation of state for the universality
class of the three-dimensional Ising model directly from Monte Carlo 
simulations. In addition we have determined the universal scaling functions
of the susceptibility and of the correlation length. There are essentially 
three difficulties one encounters in such an enterprise: finite size effects,
corrections to scaling and the 
unknown size of the critical region, where scaling works and the equation 
of state describes the critical behaviour of the magnetization. In order
to minimize the corrections to scaling we have studied the improved Ising 
model as proposed in Ref.\ \cite{Hasenbusch:1999mw}. We used cubic lattices
of linear extension $L$ in the range 24 to 120 to prevent finite size effects
and performed a large number of simulations with non-zero magnetic field
$H$ at different couplings $J=1/T$ to identify the critical region.
Further simulations at $H=0$ and at $T=T_c$ served to determine the critical
amplitudes of the magnetization, the susceptibility and the 
correlation length. In particular, we have calculated the amplitudes $B^c$
of the magnetization and $\xi^c$ of the correlation length for the first
time from MC simulations. The critical behaviours of $M$, $\chi$ and $\xi$
on these lines in the $H,T$-plane could indeed be described well without the
leading correction-to-scaling terms. From the amplitudes, which are different
from those of the standard Ising model, we have derived the universal amplitude
ratios $C^+/C^-=4.756(28)$, $R_{\chi}=1.723(13)$, $Q_c=0.326(3)$ and
$Q_2=1.201(10)$, in accord with other MC and analytic results. Only our 
estimate for $U_{\xi}=1.935(14)$ differs from the corresponding ratio for
the exponential correlation length, probably because the $\xi$-values in the 
broken phase were underestimated slightly due to the contribution of higher
states to the correlation functions.
 
As it turned out, the critical region extends only to rather small values
in the external field $H$, especially in the region below the critical
temperature. We have compared our scaling magnetization data to the
parametric representation of the equation of state by Guida and Zinn-Justin
\cite{Zinn-Justin:2001bf}-\cite{Guida:1996ep} and find excellent agreement
between theory and numerical experiment. The comparison of the susceptibility 
data to the corresponding scaling function shows a marginal difference 
in the symmetric phase, which however can be explained by the slightly
different value for $R_{\chi}$ used in the parametrization. The shape of
the correlation-length-scaling function is very similar to the one of the
susceptibility, the peak positions of the two scaling functions are 
coinciding within the error bars. The ratio $\hat g_{\xi}^2/f_{\chi}$ 
is therefore a slowly varying function without a peak, as expected from 
earlier parametrizations. 
\vskip 0.2truecm
\noindent{\Large{\bf Acknowledgments}}

\n We are grateful to Michele Caselle for enlightening discussions, in
particular about the correlation length, and the encouragement of Tereza
Mendes to start this project. We appreciate the constant interest and 
assistance of our colleagues Sven Holtmann and Thomas Schulze. Our work was
supported by the Deutsche Forschungs\-ge\-meinschaft under Grant
No.\ FOR 339/2-1.




\begin{thebibliography}{99}

\bibitem{Pelissetto:2000ek}
A.~Pelissetto and E.~Vicari,
\PRep {\bf 368} (2002) 549
[cond-mat/0012164] .

\bibitem{Zinn-Justin:1996}
J.~Zinn-Justin, {\em Quantum Field Theory and Critical Phenomena},
Clarendon Press, Oxford, 3rd Edition 1996.

\bibitem{Zinn-Justin:2001bf}
J.~Zinn-Justin,
Phys.\ Rept.\  {\bf 344} (2001) 159
[hep-th/0002136].

\bibitem{Guida:1998bx}
R.~Guida and J.~Zinn-Justin,
J.\ Phys.\ A {\bf 31} (1998) 8103
[cond-mat/9803240].

\bibitem{Guida:1996ep}
R.~Guida and J.~Zinn-Justin,
Nucl.\ Phys.\ B {\bf 489} (1997) 626
[hep-th/9610223].

\bibitem{Fisher:1999tr}
M.~E.~Fisher, S.-Y.~Zinn and P.~J.~Upton,
Phys.\ Rev.\  B {\bf 59} (1999) 14533.

\bibitem{Campostrini:1999at}
M.~Campostrini, A.~Pelissetto, P.~Rossi and E.~Vicari,
Phys.\ Rev.\ E {\bf 60} (1999) 3526
[cond-mat/9905078].

\bibitem{Engels:2000xw}
J.~Engels, S.~Holtmann, T.~Mendes and T.~Schulze,
Phys.\ Lett.\ B {\bf 492} (2000) 219
[hep-lat/0006023].

\bibitem{Toussaint:1996qr}
D.~Toussaint,
Phys.\ Rev.\ D {\bf 55} (1997) 362
[hep-lat/9607084].

\bibitem{Engels:1999wf}
J.~Engels and T.~Mendes,
Nucl.\ Phys.\ B {\bf 572} (2000) 289
[hep-lat/9911028].

\bibitem{Ballesteros:1998my}
H.~G.~Ballesteros, L.~A.~Fern\'andez, V.~Mart\'{\i}n-Mayor and 
A.~Mu\~noz Sudupe,
Phys.\ Lett.\ B {\bf 441} (1998) 330
[hep-lat/9805022].

\bibitem{Hasenbusch:1998gh}
M.~Hasenbusch, K.~Pinn and S.~Vinti,
Phys.\ Rev.\ B {\bf 59} (1999) 11471
[hep-lat/9806012].

\bibitem{Hasenbusch:1999mw}
M.~Hasenbusch,
J.\ Phys.\ A {\bf 32} (1999) 4851
[hep-lat/9902026].

\bibitem{Zinn:1996sy}
S.-Y.~Zinn, M.~E.~Fisher,
Physica A {\bf 226} (1996) 168.

\bibitem{Talapov:1996yh}
A.~L.~Talapov and H.~W.~Bl\"ote,
J.\ Phys.\ A{\bf 29} (1996) 5727
[cond-mat/9603013].

\bibitem{Privman:1991}
V.~Privman, P.~C.~Hohenberg and A.~Aharony,
in {\em Phase Transitions and Critical Phenomena}, vol. 14,
edited by C.~Domb and J.~L.~Lebowitz (Academic Press, New York, 1991).

\bibitem{Widom:1965}
B.~Widom,
J.\ Chem.\ Phys. {\bf 43} (1965) 3898.

\bibitem{Griffiths:1967}
R.~B.~Griffiths,
Phys.\ Rev.\ {\bf 158} (1967) 176.

\bibitem{Brower:mt}
R.~C.~Brower and P.~Tamayo,
Phys.\ Rev.\ Lett.\  {\bf 62} (1989) 1087.

\bibitem{Wolff:1988uh}
U.~Wolff,
Phys.\ Rev.\ Lett.\  {\bf 62} (1989) 361.

\bibitem{Dimitrovic:yd}
I.~Dimitrovic, P.~Hasenfratz, J.~Nager and F.~Niedermayer,
Nucl.\ Phys.\ B {\bf 350} (1991) 893.

\bibitem{Seniuch:2002}
M.~Seniuch,
{\em Zustandsgleichungen und Korrelationsl\"ange beim dreidimensionalen
Isingmodell mit verbesserter Wirkung},
Diploma Thesis, Universit\"at Bielefeld, March 2002.

\bibitem{Caselle:1997hs}
M.~Caselle and M.~Hasenbusch,
J.\ Phys.\ A {\bf 30} (1997) 4963
[hep-lat/9701007].

\bibitem{Engels:1999nv}
J.~Engels and T.~Scheideler,
Nucl.\ Phys.\ B {\bf 539} (1999) 557
[hep-lat/9808057].

\bibitem{Bagnuls:2001jz}
C.~Bagnuls and C.~Bervillier,
Phys.\ Rev.\ E {\bf 65} (2002) 066132
[hep-th/0112209].

\bibitem{Fisher:fs}
M.~E.~Fisher and S.~Zinn,
J.\ Phys.\ A {\bf 31} (1998) L629.

\bibitem{Campostrini:2002cf}
M.~Campostrini, A.~Pelissetto, P.~Rossi and E.~Vicari,
Phys.\ Rev.\ E {\bf 65} (2002) 066127
[cond-mat/0201180].
 
\bibitem{Schofield:1969a}
P.~Schofield, 
Phys.\ Rev.\ Lett.\ {\bf 22} (1969) 606.

\bibitem{Schofield:1969b}
P.~Schofield, J.~D.~Lister, J.~T.~Ho,
Phys.\ Rev.\ Lett.\ {\bf 23} (1969) 1098.

\bibitem{Josephson:1969}
B.~D.~Josephson, 
J.\ Phys.\ C {\bf 2} (1969) 1113.

\bibitem{Engels:1994ek}
J.~Engels, V.~K.~Mitrjushkin and T.~Neuhaus,
Nucl.\ Phys.\ B {\bf 440} (1995) 555
[hep-lat/9412003].

\bibitem{Caselle:1999tm}
M.~Caselle, M.~Hasenbusch and P.~Provero,
Nucl.\ Phys.\ B {\bf 556} (1999) 575
[hep-lat/9903011].

\bibitem{Agostini:1996xy}
V.~Agostini, G.~Carlino, M.~Caselle and M.~Hasenbusch,
Nucl.\ Phys.\ B {\bf 484} (1997) 331
[hep-lat/9607029].

\clearpage
\end{thebibliography}
\end{document}